\documentclass{aa}
\usepackage{astro_bib_macro}
\usepackage{natbib}
\usepackage{txfonts}
\bibpunct{(}{)}{;}{a}{}{,}

\usepackage{epsf}

\newcommand{\mysp}{}\def\mysp/{}
\newcommand{\allo}[3]{\ion{#1\mysp/}{#2}\ #3$\mu$m}
\newcommand{\forb}[3]{[\ion{#1\mysp/}{#2}]\ #3$\mu$m}

\newcommand{\alloa}[3]{\ion{#1\mysp/}{#2}\ #3\AA}
\newcommand{\forba}[3]{[\ion{#1\mysp/}{#2}]\ #3\AA}

\newcommand{\dforba}[4]{[\ion{#1\mysp/}{#2}]\ #3,#4\AA}

\newcommand{\ro}{} \def\ro/{r$_{[\ion{O\mysp/}{iii}]}$}
\newcommand{\hi}{}\def\hi/{\ion{H\mysp/}{i}}
\newcommand{\hii}{}\def\hii/{\ion{H\mysp/}{ii}}
\newcommand{\uchii}{}\def\uchii/{UCHII}
\newcommand{\bra}{}\def\bra/{Br$_\alpha$}
\newcommand{\masection}{}\def\masection/{\S}
\newcommand{\exar}{}{\def\exar/{[\ion{Ar}{iii/ii}]}
\newcommand{\exn}{}{\def\exn/{[\ion{N}{iii/ii}]}
\newcommand{\exs}{}{\def\exs/{[\ion{S}{iv/iii}]}
\newcommand{\exne}{}{\def\exne/{[\ion{Ne}{iii/ii}]}
\newcommand{\exall}{}{\def\exall/{\exar/, \exn/, \exs/, and \exne/}
\newcommand{\exo}{}{\def\exo/{\forba{O}{iii}{5007}/\dforba{O}{ii}{3727}{29}}
\newcommand{\heish}{}{\def\heish/{\alloa{He}{i}{5876}/H$_\beta$}
\newcommand{\teff}{}\def\teff/{$\mathrm{T}_{\mathrm{eff}}$}
\newcommand{\wmbasic}{}\def\wmbasic/{{\it WM-Basic}}
\newcommand{\costar}{}\def\costar/{{\it CoStar}}
\newcommand{\tlusty}{}\def\tlusty/{{\it TLUSTY}}
\newcommand{\cmfgen}{}\def\cmfgen/{{\it CMFGEN}}
\newcommand{\kurucz}{}\def\kurucz/{{\it Kurucz}}
\newcommand{\logg}{}\def\logg/{$\log(g)$}
\newcommand{\logu}{}\def\logu/{$\log(\bar{U})$}
\newcommand{\um}{}\def\um/{$\bar{U}$}
\newcommand{\figscode}{}\def\figscode/{Figs.~\ref{fig:comp_uv}-\ref{fig:exci1}}
\def\msun{\ifmmode M_{\odot} \else M$_{\odot}$\fi}
\def\zsun{\ifmmode Z_{\odot} \else Z$_{\odot}$\fi}
\def\lsun{\ifmmode L_{\odot} \else L$_{\odot}$\fi}
\newcommand{\kms}{km\,s$^{-1}$}
\newcommand{\dix}[1]{$\times 10^{#1}$}

\begin{document}

\title{Mid-IR observations of Galactic \hii/ regions: constraining
ionizing spectra of massive stars and the nature of the
observed excitation sequences}

\titlerunning{EUV fluxes of massive stars}
\authorrunning{C. Morisset et al.}

\author{C. Morisset \inst{1,2}
        \and D. Schaerer \inst{3,4}
        \and J.-C. Bouret \inst{2}
        \and F. Martins \inst{4,3}
    }

\offprints{C. Morisset}
\mail{Morisset@Astroscu.UNAM.mx}
\institute{
           Instituto de Astronom\'{\i}a, Universidad Nacional
        Aut\'onoma de M\'exico; Apdo. postal 70--264; Ciudad Universitaria;
        M\'exico D.F. 04510; M\'exico. 
        \and
           Laboratoire d'Astrophysique de Marseille, CNRS, BP 8,
           F-13376 Marseille Cedex 12, France
        \and
           Observatoire de Gen\`eve, 51, Ch. des Maillettes, CH-1290
        Sauverny, Switzerland 
        \and
       Laboratoire d'Astrophysique, UMR 5572, Observatoire
        Midi-Pyr\'en\'ees, 14,  Av. E. Belin, F-31400 Toulouse, France
           }

\date{\today\ post referee report}


\abstract{
Extensive photoionization model grids for single star \hii/ regions 
using a variety of recent state-of-the-art stellar atmosphere models
have been computed with the main aim of constraining/testing their
predicted ionizing spectra against recent ISO mid-IR observations of 
Galactic \hii/ regions, which probe the ionizing spectra between 
$\sim$ 24 and 41 eV thanks to Ne, Ar, and S fine structure lines.
Particular care has been paid to examining in detail the 
dependences of the nebular properties on the numerous nebular
parameters (mean ionization parameter \um/, abundances, dust etc.) which
are generally unconstrained for the objects considered here.
Provided the ionization parameter is fairly constant on average
and the atomic data is correct these comparisons show the following:
\begin{itemize}
\item
Both recent non-LTE codes including line blanketing and stellar winds
(\wmbasic/ and \cmfgen/) show a reasonable agreement with the
observations, although non-negligible differences between their
predicted ionizing spectra are found.
On the current basis none of the models can be preferred over the other.
\item
The softening of the ionizing spectra with increasing metallicity 
predicted by the \wmbasic/ models is found to be too strong.
\item
We confirm earlier indications that the \costar/ atmospheres, including
an approximate treatment of line blanketing, overpredict somewhat 
the ionizing flux at high energies.
\item
Both LTE and non-LTE plane parallel hydrostatic atmosphere codes, 
predict ionizing spectra which are too soft, especially over the energy
range between 27.6, 35.0, and 41.1~eV and above.
The inclusion of wind effects is crucial for accurate predictions
of ionizing fluxes.
\item
Interestingly, blackbodies reproduce best the observed
excitation diagrams, which indicates that the ionizing spectra of our
observed objects should have relative ionizing photon flux productions 
${\rm Q}_{\rm E}$ at energies 27.6, 35.0 and 41.1~eV close to 
that of blackbody spectra.
\end{itemize}
These conclusions are found to be fairly robust to effects such as
changes of \um/, nebular and stellar metallicity changes, and the inclusion
of dust. Uncertainties due to atomic data (especially for Ar) are discussed.
We also discuss the possibilities and difficulties in estimating absolute 
stellar temperatures from mid-IR line ratios, or softness parameters 
defined in analogy with the optical $\eta$ indicator 
$\eta_{O-S}$=(\dforba{O}{ii}{3726}{27}/\dforba{O}{iii}{4959}{5007}) /  
(\dforba{S}{ii}{6717}{31}/\dforba{S}{iii}{9069}{9532}).
\exar/ is found to be the only fairly stable \teff/ indicator.
Mid-IR $\eta$'s appear to be of limited diagnostic power both empirically and
theoretically.
Finally we have examined which parameters are chiefly responsible
for the observed mid-IR excitation sequences. The galactic gradient of
metallicity changing the shape of the stellar emission is found to be
one of the drivers for the excitation sequence of
Galactic \hii/ regions, the actual contribution of this effect being
finally atmosphere model dependent.
We find that the dispersion of \teff/ between different \hii/ regions 
due to statistical sampling of the IMF plus additional scatter in the 
ionization parameter are probably the dominant driver for the 
observed excitation scatter.
 }

\maketitle
            \section{Introduction}
\label{sec:intro}
Despite their paucity, hot massive stars are prominent contributors
to the chemical and dynamical evolution of their host galaxies.
Because of their intense nucleosynthesis, they process large amounts 
of material, on very short time scales. Furthermore, in 
addition to type II supernovae, of which they are progenitors, massive
stars drive the dynamics and energetics of the ISM through their
supersonic massive winds, thus affecting the subsequent star formation 
process in their surrounding environment.
Besides, their strong UV radiative fluxes ionize the ISM and create
H~II regions. The ionization structure of the latter is therefore,
for the most part, controlled by the EUV radiation field of their massive
stars content. 
In order to determine the properties of H~II regions, it is therefore
essential to understand the physical properties of massive stars and
most importantly, to constrain their FUV and EUV (H-ionizing continuum)
flux distribution. Yet, this part of the stellar spectrum is generally 
unaccessible to direct observations and it is crucial to find indirect 
tests to constrain it. In this context, nebular observations of H~II 
regions combined with extensive grids of photoionization models including 
state-of-the-art model atmospheres offer the best opportunity to achieve 
this goal.

In fact a large number of galactic H~II regions have
been observed with the ISO satellite \citep[see e.g][and 
references therein]{martin-H02}.
These spectra provide a wealth of spectral information,
through fine-structure lines of ions whose ionization/excitation 
threshold are located below 912 \AA. The shape of the SED in the EUV,
and more specifically the number of ionizing photons in this region,
is directly probed by ratios of successive ionization states such as
\forb{Ar}{iii}{8.98}/\forb{Ar}{ii}{6.98},
\forb{N}{iii}{57.3}/\forb{N}{ii}{121.8},
\forb{S}{iv}{10.5}/\forb{S}{iii}{18.7}, and
\forb{Ne}{iii}{15.5}/\forb{Ne}{ii}{12.8}. 
Building line ratios diagrams for these species that are very sensitive 
to different parts of the flux distribution below the Lyman threshold
allow one 
to derive informations on the actual spectral energy distribution at 
wavelengths usually unaccessible to direct observations. 
This not only provides valuable informations on the physical properties
of the H~II regions but on their stellar content as well. As a matter of
fact, it is nowadays often used to estimate the spectral type of
the ionizing source of single star H~II regions, and offers a useful 
counterparts to more classical techniques of typing, based on optical 
or near-infrared absorption features \citep{MATHYS88,HCR96,WH97,KAPER02}.  

On the other hand, modeling tools to analyze the photosphere and winds
of hot, massive stars with a high level of accuracy and reliability have
become available in the recent years. In particular, major progress has 
been achieved to model the stellar photosphere and stellar wind in an 
unified approach incorporating also a treatment of non-LTE line
blanketing for the major opacity sources \citep{HM98,PHL01,HL95,LH03,LH03e}.

The impact of the first generation of atmosphere models including
  stellar winds and non-LTE line blanketing on nebular diagnostics
  was studied by \citet{Sch97} using the \costar/ atmosphere models of
\citet{SC97}.
  This study showed already several improvements with respect to
  the widely used LTE models of \citet{K91}.
  More recently \citet{martin-H02,MH03} have investigated the metallicity
dependence of the  
spectral energy distribution of O stars and the ionization structure 
of H~II regions, using the \cmfgen/ code by \citet{HM98}. They also 
compared the EUV fluxes from \cmfgen/ to those of the \costar/ \citep{SC97}
and \wmbasic/ \citep{PHL01} codes. They concluded that different treatment 
of line-blanketing between \costar/ on the one hand and \wmbasic/ and \cmfgen/ 
on the other hand results in significant differences in the predicted 
EUV SEDs and ionizing fluxes. 

In this context, it is of special interest to investigate how the different
models available nowadays, compare to each other in predicting nebular lines 
ratios. Similarly, it is of importance to test the role that a handful of 
various nebular parameters might have on the line ratios diagrams provided 
by ISO observations. 
The parameters influencing the ionization structure of a photoionized
region are: 
1) the geometry, the density distribution, the metallicity of the gas, 
and the possible absorption of the ionizing radiation by dust, 
2) any physical quantity affecting the shape of the ionizing flux like, 
for example, the effective temperature of the ionizing star, its metallicity, 
the presence of a wind and the characteristics of the latter, 
3) the hypothesis used to model the atmosphere like the number 
of elements taken into account, the treatment of the line-blanketing,
etc... in summary: the physical ingredients and the related
assumptions used to model the emitting atmosphere.

The present paper describes photoionization models performed with
various atmosphere models, separating the effects of all these
parameters.
The paper is structured as follows:
The various adopted atmosphere models are briefly described and
compared in Sect. \ref{sec:stel_atm}.
The ionizing spectra from these models are then used as input to our
photoionization code for the calculation of extended grids
of nebular models (Sect.~\ref{sec:grid}).
The compilation of ISO observations of \hii/ regions is described
in Sect.~\ref{sec:isouchii}.
In Sect.~\ref{sec:results} we compare our photoionization models
to the observations and discuss the effect of changing parameters one
by one on the excitation diagnostics.
In Sect.~\ref{sec:diag} we test the validity of the different
excitation diagnostics and softness radiation parameters for the
determination of \teff/.
The discussion takes place in Sect.~\ref{sec:disc}
Our main conclusions are summarized in Sect.~\ref{sec:concl}.

            \section{O star atmosphere models}
\begin{figure}
\epsfxsize=9.0cm  \epsfbox{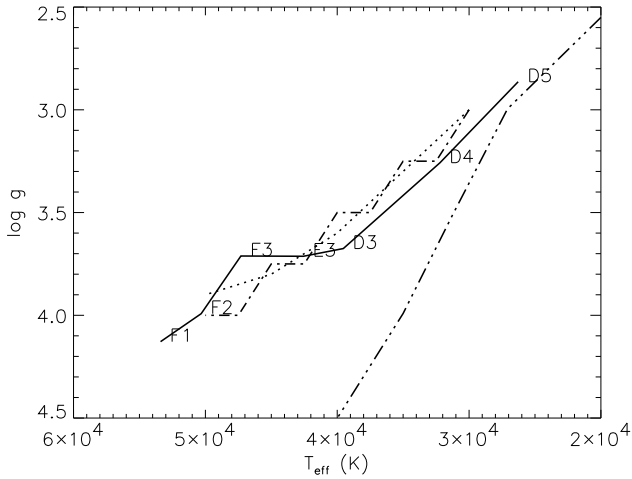}
\caption{Position in a \logg/ versus \teff/ diagram of the Supergiant
models and the \kurucz/ dwarf models
used in this paper. 
See Fig.~\ref{fig:comp_uv} for the line symbols.
\costar/ models are labeled according to the original naming convention.
\cmfgen/ models have the same parameters than \wmbasic/
models and are not drawn here.  \label{fig:logg_teff}} 
\end{figure}

\label{sec:stel_atm}
Among the key ingredients for the description of O star model atmospheres
are the treatment of non-LTE effects, the inclusion of stellar winds,
and a treatment of line blanketing \citep[see e.g.][]{AH85,KYF88,GGKP89}.
In recent years considerable improvements have been made in the
modeling of these processes and model grids computed with various 
sophisticated atmosphere codes have become available
\citep[see e.g.\ the recent conference on ``Stellar atmosphere
modeling'',][]{Hub03}.
For our photoionization models, we adopt the ionizing spectra predicted 
from five different codes (\kurucz/, \tlusty/, \costar/, \wmbasic/, \cmfgen/)
briefly described hereafter.
With the exception of the \tlusty/ and \kurucz/ models, which assume a plane
parallel 
geometry and thus no wind, all models describe the photosphere and
winds in spherical geometry, in a unified manner.  

Except mentioned otherwise, we have used atmosphere models computed
for solar abundances: He, C, N, O, Ne, Si, S, Ar and Fe being 0.1,
4.7\dix{-4}, 9.8\dix{-5}, 8.3\dix{-4}, 5.4\dix{-5}, 4\dix{-5}, 1.6\dix{-5}, 6.8\dix{-6} and 4\dix{-5} resp. 

\subsection{\kurucz/ models}
\label{subsec:kurucz}

We use the well-known plane parallel LTE line blanketed models of
\citet{K91,KU94}. Computations were done for models with \teff/ (and
\logg/) between 26 and 50~kK (3.0 and 5.0). For stability
reasons, the available 
high \teff/ models are restricted to cases of high gravity.
The employed \kurucz/ models are therefore representative of dwarfs
rather than supergiants mostly considered for the other model
atmospheres (cf.\ below).

\subsection{\tlusty/}
\label{subsec:tlusty}
A grid of plane-parallel non-LTE line blanketed models based on the \tlusty/ 
code of \citet{HL95} has recently been calculated using a super-level
approach and an Opacity distribution function or a modified opacity
sampling \citep{LH03,LH03e}.
About 100,000 individual atomic levels have been included, for more than
40 ions of \ion{H}, \ion{He}, \ion{C}, \ion{N}, \ion{O}, \ion{Ne}, \ion{Si}, 
\ion{P}, \ion{S}, \ion{Fe} and \ion{Ni}, using a superlevel approach.
For all the models, a standard microturbulent velocity 
V$_{\rm turb}$=10 \kms\ has been used.
The parameter space of the grid covers 
27\,500\,K $\leq$ \teff/ $\leq$ 55\,000\,K 
with 2500 K steps and 3.0 $\leq$ \logg/ $\leq$ 4.75 with 0.25 dex steps.
Up to 10 different metallicities, from 2 times solar to metal free chemical
composition, have been considered by \citet{LH03,LH03e}\footnote{This
grid is available at 
http://tlusty.gsfc.nasa.gov.}. We extracted from this database models with
\teff/ (\logg/)  ranging from 30 to 50~kK (3.0 to 4.0), with solar
metallicity.

\subsection{\costar/}
\label{subsec:costar}
The \costar/ models of \citet{SC97}
include stellar winds, treat H-He in full non-LTE, and include 
line blanketing effects with an opacity sampling method based 
on Monte-Carlo simulations \citep{SCHMUTZ91}.
The impact of these effects on the ionizing fluxes and nebular diagnostics
of \hii/ regions has been discussed in detail by \citet{Sch97}.

For our computations we use \costar/ models with the lowest value for
\logg/, i.e. models D5, D4, D3, E3, F3, F2 and F1 from the 
\costar/ grid of \citet{SC97}. The \teff/ (and \logg/)
range from 27~kK (2.9) to 53~kK (4.1).

\begin{figure*}
\epsfxsize=18.cm  \epsfbox{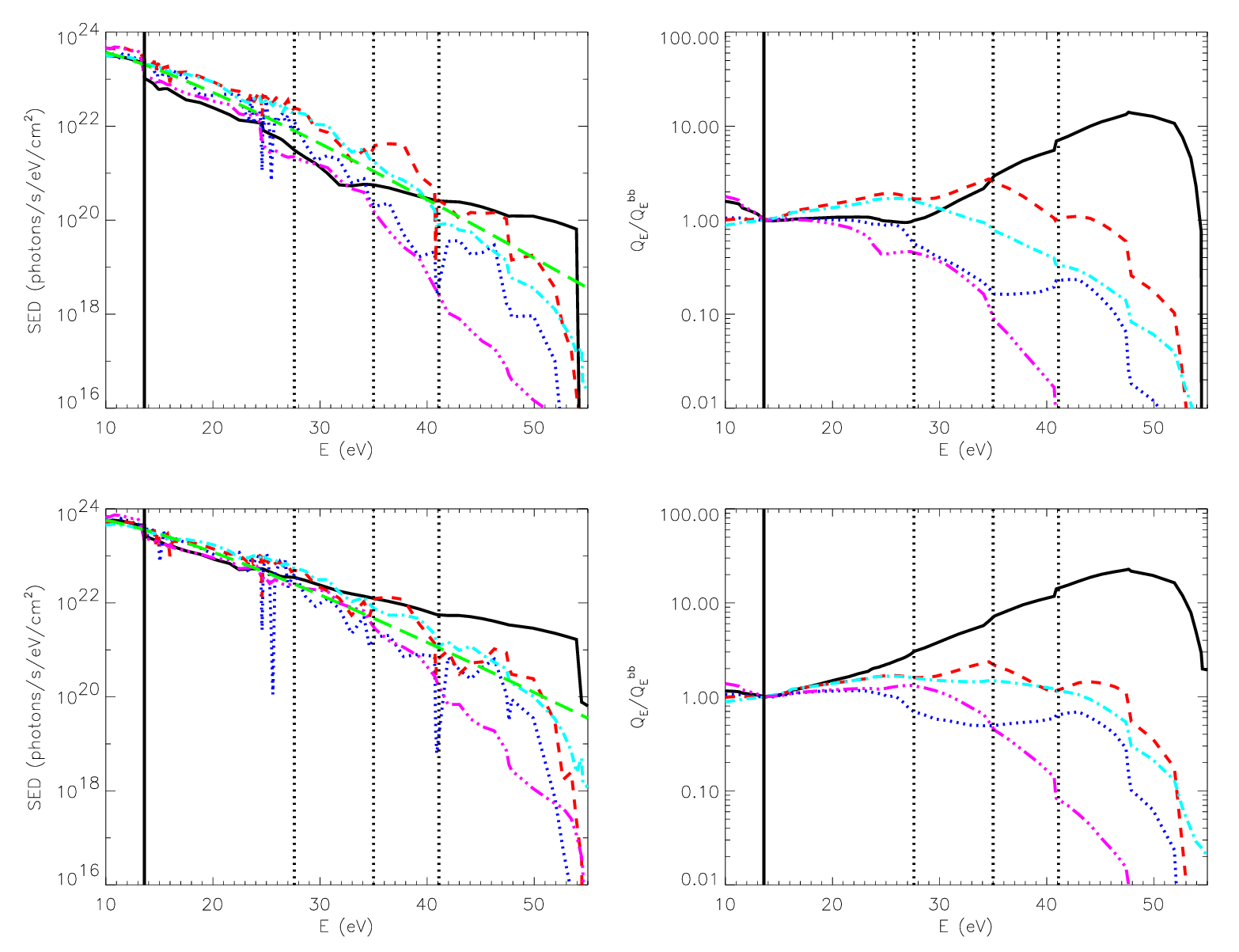}
\caption{Comparison between the 6 stellar atmosphere models:
\costar/ (solid), \wmbasic/ (dotted),
\cmfgen/ (dashed), \tlusty/ (dash dot), \kurucz/ (dash dot dot) and
the Blackbody (long dashes, left panels only), for the same 
\teff/ of 
35~kK (upper plots), except for \costar/ model (see text), 
and 40~kK (lower plots). The left
panels show the Spectral Energy Distribution and the right panels
show, for any energy E(eV), the number Q$_{\rm E}$ of photons with
energy greater than E, relative to the corresponding
number for the Blackbody emission, all the spectra having the same
value for Q$_{13.6}$. Vertical lines are plotted at 13.6~eV (solid)
and 27.6, 35.0 and 41.1~eV (dotted), corresponding to the ionization
potentials of the ions considered in this paper (Ar$^+$, S$^{++}$, and
Ne$^+$ resp.).\label{fig:comp_uv}} 
\end{figure*}

\subsection{\wmbasic/}
\label{subsec:wmbasic}
The \wmbasic/ models of \citet{PHL01} treat a large number
of ions in non-LTE and include their line blocking effect by means of an
opacity sampling technique.
The atmospheric structure is computed from the hydrodynamic equations
including radiative acceleration from numerous metal-lines and continua.
We used the grid available on the 
web\footnote{http://www.usm.uni-muenchen.de/people/adi/Models/} and
described in \citet{PHL01} for Supergiant models with \teff/
(and \logg/) ranging from 30~kK (3.0) to 50~kK (3.9). 

\subsection{\cmfgen/}
\label{subsec:cmfgen}
We have constructed spherically symmetric wind models, using the non-LTE,
comoving frame code \cmfgen/ \citep{HM98}. 
This code solves the radiative transfer equation, together with the statistical
equilibrium equations, and line blanketing is self-consistently taken into 
account,
using a super-level formulation. The chemical elements included in our model
calculations are \ion{H}, \ion{He}, \ion{C}, \ion{N}, \ion{O}, 
\ion{S}, \ion{Si}\ and \ion{Fe}. 
For the 28 ions explicitly treated, a total of 2292 levels 
distributed in 819 superlevels 
are included, representing 22762 bound-bound transitions. 
Atomic data for \ion{Fe}{iv} and \ion{Fe}{V} have been slightly
improved, compared to those  
first introduced in \cmfgen/ \citep{HM98} and made consistent with those 
used in \tlusty/. 
As shown in \cite{JCB03}, this was required to get a very good agreement 
in the determination of iron abundances, when fitting lines of these two 
successive ionization stages in individual O stars in NGC~346, 
the largest H~II region in the SMC.
The temperature structure is calculated from the assumption of radiative 
equilibrium. 
The atmospheric structure consists of the wind, parameterized by the
classical $\beta$-law, which is connected to hydrostatic layers
obtained from the ISA-Wind code of \citet{dK96}, such that at the connecting
point both the velocity and velocity gradient match.
We assume a constant Doppler profile of 20~\kms\ for all lines. 
As shown by \citet{M02} for dwarfs and by additional test calculations
this assumption leads to negligible changes of 
emergent spectrum.  
The stellar parameters, including the abundances, used to compute the \cmfgen/
grid of supergiants are identical to those used by \wmbasic/ and
described in \citet{PHL01}.

\subsection{Atmosphere models for Dwarf stars}
\label{sub:dwarfsdescr}
The main results of the present paper are obtained for Supergiant
stars. Nevertheless, we also computed grids of photoionization models
using Dwarf stellar atmosphere models as ionizing spectrum, to check
the effect of \logg/ on the excitation of the nebula. In this purpose, 
\wmbasic/ D models from \citet{PHL01} have been used. 
We have also computed \cmfgen/ models using the same set of
parameters than those used for the \wmbasic/ D models.

The models using Dwarf stellar atmosphere are discussed in 
Sect.~\ref{sub:dwarfs}.

\subsection{Rebinning of the ionizing spectra}
\label{sub:rebin}

For subsequent use in our photoionization code NEBU (described in
Sect.~\ref{sec:grid})
the different atmosphere models have to be rebinned to the energy grid 
used in this code. The SEDs are first converted to
the units used in NEBU (number of 
photons/eV/s/cm$^2$).  
The SED is then interpolated on the NEBU grid, such as to preserve
the integrated number of photons in each energy interval.
For most of
the energy intervals, the number of points in the original
stellar atmosphere domain is some tens to some hundreds, giving a good
accuracy for the  rebinning. Note that despite the low number of
points (144) used to 
describe the ionizing spectrum in NEBU, the results are
reliable, as the most important quantities are the number of photons
able to ionize the different ions. 
The grid points actually correspond to the
ionization potential of all the ions taken into account in the
photoionization computation.

\subsection{Comparing the EUV spectra}
\label{sub:comp_uv}

Fig.~\ref{fig:logg_teff} present all the Supergiant models used in
this paper in a \logg/ versus \teff/ diagram. The values for \logg/ at a
given \teff/ are very close together,
with the exception of the \kurucz/ models, which have a systematic higher
value for \logg/, up to be even higher than the value adopted for
Dwarf models \citep[see also Fig.~1 in ][]{SC97}.

Fig.~\ref{fig:comp_uv} illustrates the differences in the SED obtained
from different atmosphere models 
after the rebinning procedure described above, 
for the same \teff/, here 35 and 40~kK, with the exception of \costar/
model for which no value at 35~kK is available in the Supergiant
subset of models used here, model D4 at 32.2~kK is plotted. 
While the five models agree quite well in the domain of
energies lower than 20~eV (and very well in the optical and IR range,
not shown here), their differences can be as big as 4 orders of
magnitudes just before 4 Rydberg. In this paper, we will use IR lines
to trace the SED between 27, 35 and 41~eV (see next Section), where the
models differences already reach 1 to 2 orders of magnitude. 

Of more interest for the analysis of the behavior of the
fine-structure lines is the distribution of the ionizing photons at
each energy. This is quantified by Q$_{\rm E}$, which is the number of
photons with energy 
greater than E, shown in right panels of Fig.~\ref{fig:comp_uv}.
More precisely, the relevant quantity determining the nebular structure
and properties would be the photon output weighted by the photoionization 
cross section. 
In the range
of energy traced by the excitation diagnostics, 27-41~eV, the
behavior of the four models is very different. We will discuss this
further in Sect. \ref{sub:scomp_obs} and \ref{sub:teff}.

            \section{Grid of photoionization models}
\label{sec:grid}

Extensive grids of photoionization models were computed with the NEBU code
\citep{MP96,P02,PaperIII} in order to evaluate in detail the dependence 
of the mid-IR atomic fine-structure line emission of Galactic \hii/ regions on 
the atmosphere models, and the stellar and nebular properties.
Our main aims are a) to derive constraints on the 
stellar ionizing spectra and b) to examine the origin of the observed
excitation gradients in (compact) Galactic \hii/ regions.

In principle nebular emission line properties depend on fairly a large number
of parameters, namely:
\begin{itemize}
\item the shape of the stellar ionizing spectrum, determined (mostly)
by the stellar 
        temperature \teff/, gravity, and metallicity $Z_\star$;
\item the ionization parameter $U(r)= Q_{13.6}/N_e 4\pi r^2c$. 
As the geometry of the \hii/ regions
modeled here is an empty cavity surrounded by an ionized shell, we
prefer to use hereafter the mean ionization parameter
$\bar{U}=U(\bar{r})$, computed following
\citet{ED85} at a distance from the ionizing star $\bar{r} =
r_{empty}$ + $\Delta$R/2, where $r_{empty}$ is the size of the empty
cavity and $\Delta$R the thickness of the \hii/ shell\footnote{For $U$
derived at the Str\"omgren radius without empty cavity 
one has $U \propto  
  (Q N_e \epsilon^2)^{1/3}$, with the filling factor $\epsilon$.}. \um/ is
  essentially given by the ratio of the ionizing photon density over the 
  nebular particle density, i.e.\ properties of the ionizing source and the
  nebular geometry; 
\item the nebular abundances/metallicity $Z_{\rm gas}$
\item atomic parameters driving the ionization equilibrium and
line emissivities, e.g. photoionization cross section, recombination
coefficients (radiative and dielectronic), collisional excitation
cross sections, etc.
\item other secondary parameters like the presence of dust.
\end{itemize}
There is no doubt on the existence of a systematic metallicity variation
among the observed sources considered below. 
On the other hand, as for most of these objects (compact/ultra-compact \hii/ 
regions) the properties of the ionizing source(s) and their geometry are 
not known, it is imperative to assess the impact of all parameters on the 
observables and to establish that the conclusions drawn from comparisons 
with observations are robust in this respect.
The dependence of the emission line properties on the above parameters
is examined in Sect.~\ref{sec:exci} with the help of photoionization models 
computed for a wide range of model parameters. 

The bulk of ``standard'' models were computed for the following
cases/assumptions.
The ionizing spectra from the five atmosphere models described in
Sect. \ref{sec:stel_atm} and plotted in Fig.~\ref{fig:comp_uv}
plus blackbody spectra are adopted.
Stellar \teff/ ranging from 30 to 50~kK were used. This range 
in  \teff/ is likely to describe the physical conditions of the sample of
\hii/ regions \citep{M03}. 
For most cases we assume a solar composition for the nebular
and stellar abundances. 
Metallicity variations are considered in Sect.~\ref{sub:metal}.
For each of these stellar atmosphere, series of photoionization models
were computed for the following nebular parameters.
We set the electron density to 10$^{3}$ cm$^{-3}$, one order of magnitude
below the lowest critical density of the lines under consideration
\citep[cf.][]{PaperII}.
An empty cavity of radius 3\dix{17}cm is assumed. The
luminosity of the ionizing star is adjusted to lead to a constant
number of Lyman continuum photons (${\rm Q}_{13.6}=1.5 \, 10^{49}$ s$^{-1}$)
corresponding to an ionization parameter \logu/ = -1.5.
Additional models quantifying the effect of variations of \um/ 
are presented in Sects.~\ref{sub:maineffects}, \ref{sub:u}, and
\ref{sub:eta}. 

The effect of dust can be included in the
photoionization computation, with two different optical properties
corresponding to graphite or astronomical silicate (see Sect.~\ref{sub:dust}).

The observables predicted from these extensive model grids
will be compared to observations in Sect.~\ref{sub:scomp_obs}.

            \section{The ISO \hii/ regions catalogs}
\label{sec:isouchii}

Infra-red spectra between 2.3 and 196~$\mu$m were taken from a sample
of 43 compact \hii/ regions using the two spectrometers (SWS and LWS)
on board ISO \citep{PaperI}. Details about the data reduction and a first 
analysis of
the ionic lines in terms of abundances can be found in \citet{PaperII}. 
Error bars on the lines intensities are within 10\% to 20\%.
Note that a detailed study of one source has been achieved in \citet{PaperIII}.

\citet{GSL02} published a catalog of 112 \hii/ regions observed by ISO
SWS spectrometer. Some of the sources are common with the
\citet{PaperII} catalog. The two catalogs are very coherent in terms
of line intensities, as concluded by \cite{GMS02},
and are therefore included in our analysis.
The effect of local and interstellar attenuation, even if lower in the 
IR range used in this work than for the optical domain, can be
important and need to be corrected for. We correct the observed line
intensities from the reddening using the extinction law described in
Tab. 2 of \citet{GMS02}.

In the SWS and LWS spectral domain, 4 fine-structure line ratios are
sensitive to the ionizing flux 
distribution: \forb{Ar}{iii}{8.98}/\forb{Ar}{ii}{6.98},
\forb{N}{iii}{57.3}/\forb{N}{ii}{121.8},
\forb{S}{iv}{10.5}/\forb{S}{iii}{18.7}, and
\forb{Ne}{iii}{15.5}/\forb{Ne}{ii}{12.8}, hereafter \exall/
respectively. 

The excitation ratio implying Nitrogen lines will not be used in the
next discussion, since: 1) the two Nitrogen lines are observed by LWS
spectrometer, with a larger aperture size than the SWS: some Nitrogen
emission can arise from regions not seen in the other lines; 2) \citet{GSL02}
have observations only with SWS and then the number of observational
constraints strongly decrease when using only \citet{PaperII} data; 3) the
critical densities of the Nitrogen lines are very low compared to the one of
the other lines \citep[see][]{PaperII} and will not be emitted by
medium density gas which can still
emit the other lines, and 4) the ionization potential of N$^{++}$ is
very close to the one of Ar$^{++}$ (29.7 and 27.6~eV resp.), so the
main conclusions regarding the 30~eV energy domain will be obtained
from Argon lines. 

Depending on the element, the number of
sources for which we have finite value for the corrected excitation
ratios is ranging from 45 to 51. Error bars on the line intensities
are approximately 10 to 20\%.

            \section{Excitation diagnostics}
\label{sec:exci}
\label{sec:results}

\begin{figure}
\epsfxsize=9.0cm  \epsfbox{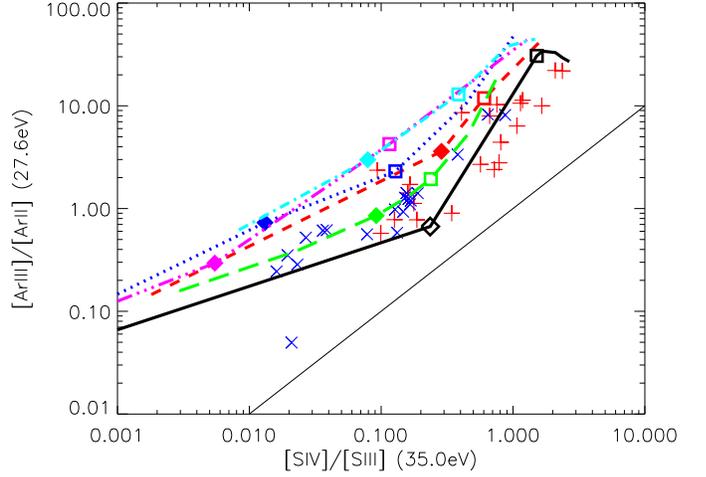}
\caption{Deredenned observed values for the excitation sensitive line
 ratio \exar/ versus \exs/ (the corresponding ionization potentials are also
 given). Source with a galactocentric distance lower than 7~kpc are
 symbolized with a +, otherwise with an X.
 Results from the photoionization model grid are line plotted using the 
 same codes as in Fig.~\ref{fig:comp_uv}. The
 plot have been done such as the lowest ionization potential (indicated
 in braces) is always on the y-axis. Models obtained with 35 and 40~kK
 stars are shown using filled diamonds and empty squares respectively
 (except for \costar/ model at 32.2~kK, empty diamond, see text). 
 The y=x line is also drawn.
 \label{fig:exci1}} 
\end{figure}
\begin{figure}
\epsfxsize=9.0cm  \epsfbox{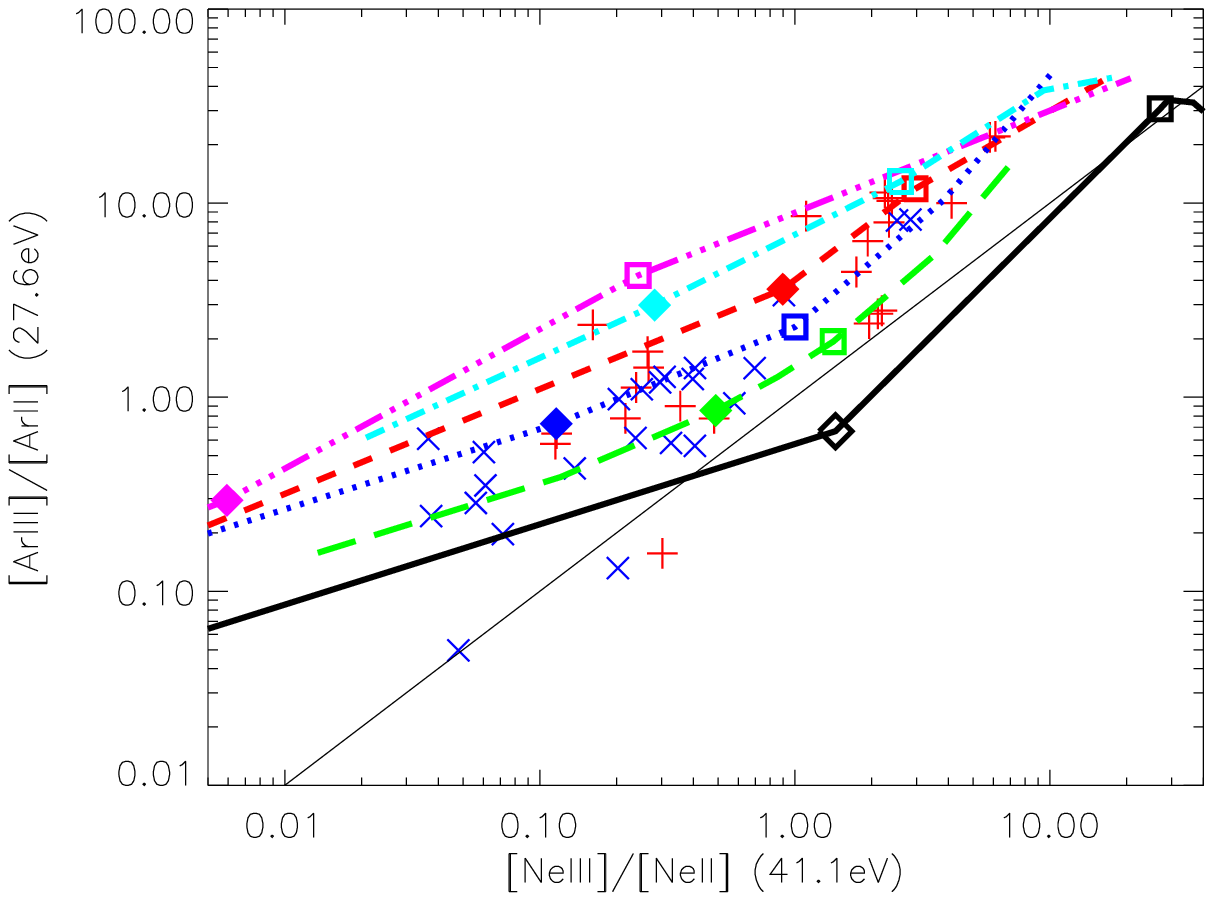}
\caption{Same as Fig.~\ref{fig:exci1} for the excitation sensitive line
 ratio \exar/ versus \exne/
 \label{fig:exci2}} 
\end{figure}
\begin{figure}
\epsfxsize=9.0cm  \epsfbox{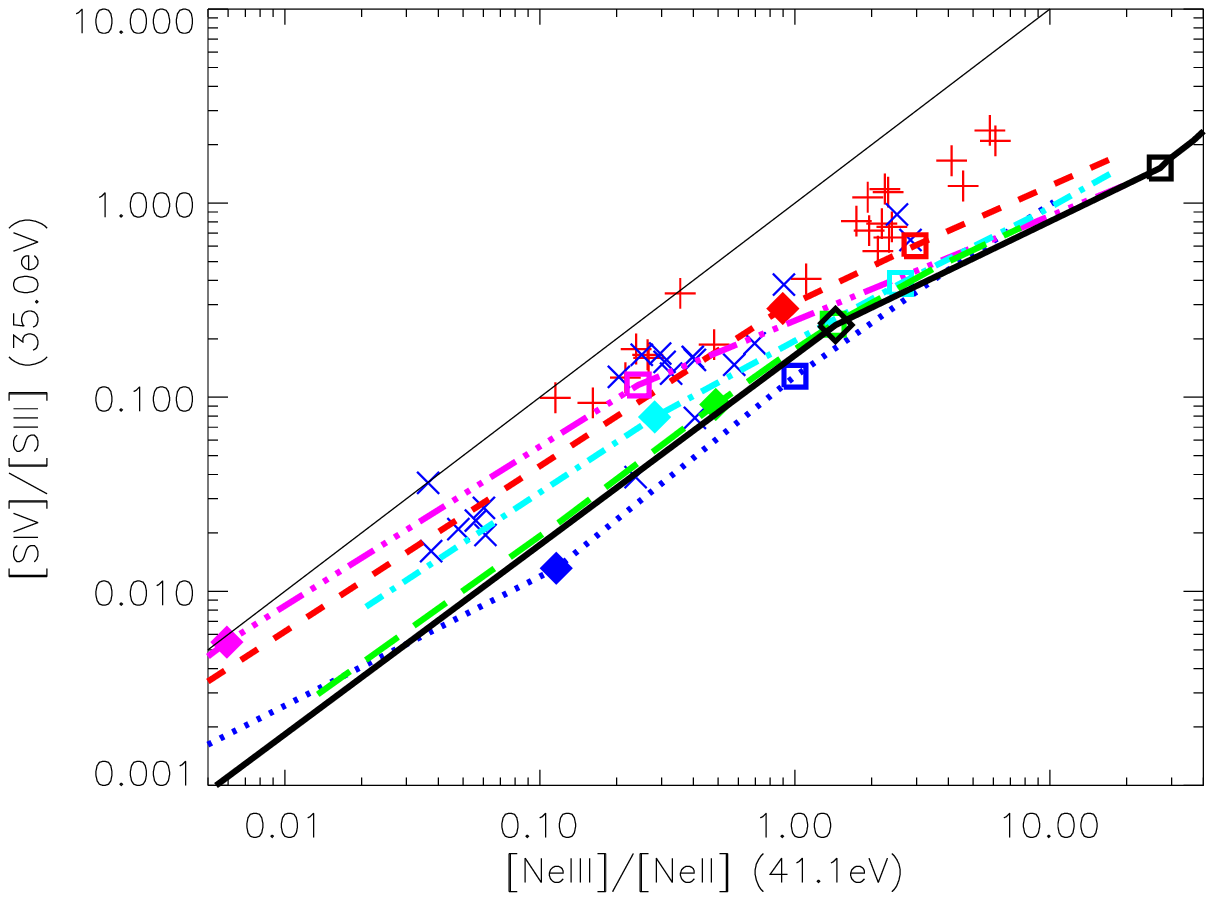}
\caption{Same as Fig.~\ref{fig:exci1} for the excitation sensitive line
 ratio \exs/ versus \exne/
 \label{fig:exci3}} 
\end{figure}

Figs. \ref{fig:exci1} to \ref{fig:exci3} show the main results of the
photoionization models using different 
atmosphere models for the ionizing star (lines), and the deredenned observed
values,
for the two merged catalogs \citep{PaperII,GSL02}. As we consider 3
diagnostic ratios, 3 plots can be drawn. The models obtained with
\teff/~=~35 and 40~kK are symbolized by filled diamonds and open squares
respectively. The open diamond indicates a \costar/ model at 32.2~kK
as no model at 35 kK is available.

In principle the position of a model in such diagrams depend on all
the following parameters:
the hardness of the stellar SED (parametrized here for each set of model 
atmospheres by \teff/ for a fixed stellar metallicity) 
and the main nebular parameters, i.e.\ the ionization parameter \um/ 
and the nebular composition.
Let us consider first the case of constant (solar) metallicity and
constant \um/ (but see Sects.~\ref{sub:maineffects}, \ref{sub:u} 
and \ref{sub:metal}).
In this case the location of a point on such diagnostic 
diagrams depends only on {\em a)} the global excitation of the gas 
and {\em b)} the ``slope'' of the ionizing photon distribution between the two
corresponding ionization potentials.
For constant \um/ and a given set of atmosphere models the excitation 
({\em a}) is determined by \teff/. In other words, when \teff/ increases, 
the number of ionizing 
photons at all the energies traced by the observed ions increases,
and the points in the excitation diagrams will essentially move 
along the diagonal ($y=x$) direction.
Note, however, that different atmosphere models with the same \teff/
predict fairly different absolute positions in these plots. This
simply reflects the differences in the predicted number of ionizing 
photons above the relevant energy (cf.\ Fig.~\ref{fig:comp_uv}).

For a given line ratio the other line ratios depend to first order
on the ``slope'' of the ionizing spectra ({\em b}).
More precisely, the relevant quantity is the slope of 
the cumulative number of ionizing photon flux ${\rm Q}_{\rm E}$ between the
corresponding ionization potentials (see right panels in 
Fig.~\ref{fig:comp_uv}).
For example, \tlusty/ and \kurucz/ models show in general the softest spectra
(i.e.\ steepest slopes) between 27.6 and 41.1~eV. 
For a given \exne/ these models therefore show the highest \exar/ values.

For the assumptions made here (constant \um/ and metallicity)
every location of the model results in
the three excitation diagrams can be approximately understood in terms of 
the ionizing photon distributions ${\rm Q}_{\rm E}$. 
The correspondence is not always exact, as
some competitive processes take place in the use of the ionizing
photons, but the overall trends can be simply understood from
the shapes of the spectra.
Other additional assumptions (e.g.\ on the luminosity class of the 
exciting sources, the presence of dust, and uncertainties of the
atomic data) also affect the predicted excitation diagrams.
These effects are discussed below.

The observed excitations, correlated between the three
excitation ratios \exar/, \exs/, and \exne/, can be decomposed into two 
components: an {\bf excitation sequence} showing a global increase of
the excitation ratios over $\sim 2$ orders of magnitude, following to
first order a trend parallel to $y=x$ in the excitation diagrams, and a 
{\bf superposed excitation scatter} of typically $\sim 0.5-1$ order of
magnitude around the mean excitation 
(cf.\ Figs.~\ref{fig:exci1} to \ref{fig:exci3}, 
but see also Figs.~\ref{fig:exar_z} and
\ref{fig:exne_z}, where excitations versus 
metallicity are plotted).

\subsection{First comparison with the observations}
\label{sub:scomp_obs}

The zero-th order trend of the
observations plotted in Figs.~\ref{fig:exci1} to \ref{fig:exci3}
is reproduced by the models: the excitation of the ionized gas,  
traced by the X$^{i+1}$/X$^i$ ratios, are well correlated. 
A \teff/ sequence from $\sim$ 30 to 45~kK succeeds in
reproducing the entire range of gas excitations.
Note however, that, as discussed below (Sect.\ \ref{sub:origin}),
this does not imply that the ionizing stars of our objects 
indeed cover this range of \teff/.

Fairly large differences are found in the predicted excitation diagnostic
diagrams (Figs.~\ref{fig:exci1} to \ref{fig:exci3})
when using different atmosphere models.
As expected from the intrinsic SEDs, 
the largest differences are found in Fig.~\ref{fig:exci2}, which
traces the largest energy domain ($\sim$ 28 to 41~eV) corresponding
to the \exar/ and \exne/ ratios.

When taken literally (i.e.\ assuming a fixed constant value of \um/ 
for all atmosphere model sets and a fixed solar metallicity) 
Figs.~\ref{fig:exci1} to \ref{fig:exci3} indicate the following
concerning the shape of the ionizing spectra.
\begin{enumerate}
\item
Both recent codes, \wmbasic/ and \cmfgen/, show a reasonable
agreement with the observations.
Given their different behavior in the three excitation diagnostics
depending on luminosity class
(cf.\ Sect.\ \ref{sub:dwarfs}), and other remaining uncertainties 
discussed subsequently, it seems quite clear that none of the models
can be preferred on this basis.

Despite these similarities we note, however, that an important
offset is found in the excitation ratios predicted by
these codes for a given absolute value of \teff/ 
(cf.\ Sect.\ \ref{sub:teff}).
\item 
Both plane parallel hydrostatic codes (\kurucz/, \tlusty/)
predict spectra which are too soft, especially over the energy
range between 27.6 and 41.1~eV and above.
For the \kurucz/ LTE models this problem is just a manifestation
of the well-known ``Ne~{\sc iii}'' problem \citep{R88,SCR95,Sch97}.
This problem is persistently found with the non-LTE \tlusty/ models.
Although not completely clear, 
this ``softness'' is likely due the neglect of wind effects
which are known to alter the ionizing spectrum \citep[cf.][]{GGKP89,Sch97}
albeit in fairly complex way involving line blanketing
from large numbers of metal-lines.

\item
As already found in other investigations \citep[cf.][and references
therein]{ODSS00,S00} we see that the \costar/ models  
(showing the hardest spectra among the ones considered here) 
overpredict somewhat the ionizing flux at high energies.
This likely overestimate of the ionizing
flux at high energy (cf.\ below) must be due to the approximate
and incomplete treatment of line blanketing \citep[see also][]{CRO99,MH03}.

\item Interestingly blackbodies reproduce best the observed
excitation diagrams, which indicates {\em a posteriori} that the
ionizing spectra should have relative ionizing
photon flux productions (${\rm Q}_{\rm E}$ at energies 27.6, 35.0 and 
41.1~eV) close to that of blackbody spectra.
This will be discussed in more detail in Sect.\ \ref{sec:disc}.
\end{enumerate}

We note that the results concerning \cmfgen/, \wmbasic/,
\costar/, and \kurucz/ models presented here confirm and support those
presented 
by \citet{SSZ02} in terms of position of the models between each other
in the \exar/ versus \exne/ excitation diagram and distance to the \hii/
regions observations (cf.\ below).

\subsection{Main dependences: stellar temperature, ionization parameter,
metallicity}
\label{sub:maineffects}

For clarity it is useful to discuss first the dependences of the 
excitation diagnostics on the main parameters, i.e.\ the stellar temperature, 
ionization parameter, and metallicity. 
This is illustrated here somewhat schematically for the case of the 
\exne/ versus \exar/ diagnostic. Qualitatively the same results are found for
the other excitation diagrams.

\begin{figure}
\epsfxsize=9.0cm  \epsfbox{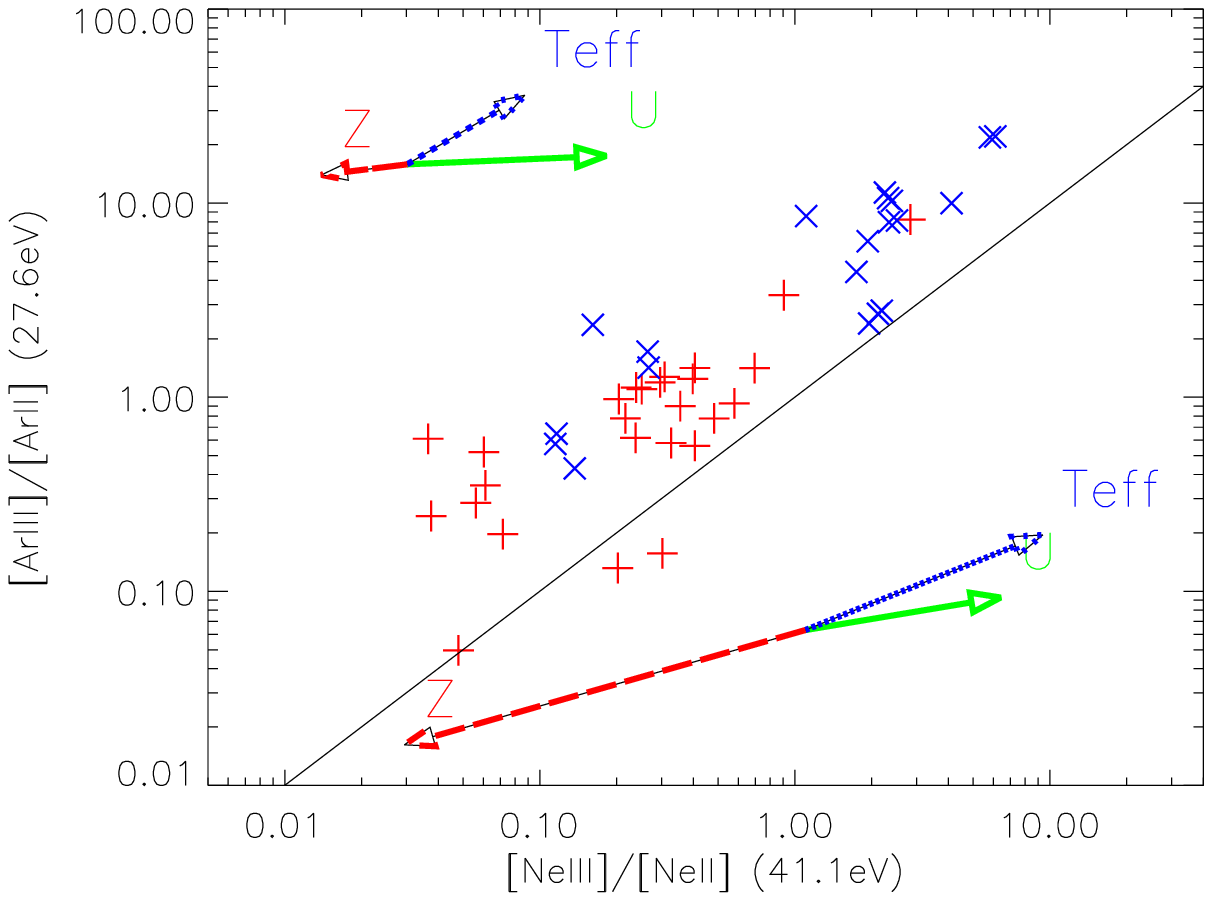}
\caption{Increase of the excitation diagnostics \exar/ versus \exne/,
when \logu/ is increased from -2.6 to -0.8 (solid arrows), when the
metallicity of 
the gas is increased from half 
solar to twice solar (dashed arrows), and when \teff/ is increasing
from 35 to 40~kK (dotted arrows). Upper set of arrows are for
BlackBody models, and 
lower set for \wmbasic/ models. In the latter case the metallicity is
changes coherently for both the gas and the star. Codes are the same
than in \figscode/. 
\label{fig:maineffects}\label{fig:u_and_metal}}
\end{figure}

An increase of the stellar temperature \teff/ or ionization parameter \um/ or
a decrease of the metallicity all lead overall to a higher excitation of 
the nebula, which e.g.\ manifests itself by larger \exne/ and \exar/ line 
ratios. However, although both line ratios change
in similar ways, their effect is distinguishable to some extent.
This is illustrated in Fig.~\ref{fig:maineffects}, which shows
for blackbody (and \wmbasic/, see  Sect.~\ref{sub:metal}) 
spectra the implied shift in the \exne/
versus \exar/ 
excitation diagnostics due to a change of \teff/, \um/ and $Z_{\rm gas}$
(consistent changes of both the stellar and nebular metallicity are 
discussed in Sect.~\ref{sub:metal}).
From this figure we see that \teff/ and \um/ variations are not completely
degenerate (i.e.\ ``parallel''). 

Also, an increase of the nebular metallicity leads to a (very) small
decrease of the excitation diagnostics, quite parallel to the
variations induced by changing \um/. The effect of changing
coherently the metallicity of both the \hii/ region and the star
(which is principally acting on excitation diagnostics) are
considered in Sect.~\ref{sub:metal}.

The effect of continuum absorption by dust inside the \hii/ region 
on the excitation diagnostics is also parallel to changes of \teff/, 
and is discussed in Sect.\ \ref{sub:dust}.
Although quantitatively these variations depend e.g.\ on the adopted
SED (and of the point in the \teff/-\um/-Z space chosen to compute
the partial derivatives traced by the arrows in
Fig.~\ref{fig:maineffects}), these qualitative distinctions remain
valid for the  
entire parameter space considered in the present paper and will be
useful for the discussions below.

Note that earlier investigations have considered that changes of \um/
are completely degenerate with 
\teff/ \citep{GSL02,MVT02}\footnote{For their grids the ionization
parameter $U$ defined at the  
Str\"omgren radius vary actually by $\sim$ 0.1 (supergiant grid) and
0.3 dex (dwarfs).}, 
or have simply assumed an arbitrarily fixed, constant value of \um/ 
\citep{MH03}.

As apparent from Fig.~\ref{fig:maineffects}, \exar/ varies little with
\um/ in comparison with other mid-IR excitation ratios. This behavior is
easily understood, as explained in the following brief digression.

\label{sub:heish}
In \teff/ and \um/ domain used in this
work, the \exar/ ratio is controlled by the Helium equilibrium: the
IPs of Ar$^{+}$ and He$^0$ are closed together (27.7 and 24.6~eV
resp.); in this energy domain the photon dominant predator is
He$^0$. The Ar$^{++}$ region (Ar$^{+}$) is then 
cospatial with the
He$^+$ (He$^0$) region (Some Ar$^+$ can also be present in the He$^+$
region, depending on the ionization parameter). 
The \hii/ regions modeled here are all
radiation bounded, the size (and the emission) of the He$^+$ and
Ar$^{++}$ region is mainly proportional to Q$_{24.6}$, while the size
of the He$^0$ and Ar$^{+}$ region
is controlled by the size of the \hii/ region removing the He$^+$ region. 
\exar/ is then mainly controlled by
Q$_{24.6}$/Q$_{13.6}$. The previous argumentation is valid only
if the recombination of Ar$^{++}$ remains quite small, which is not
the case when strong dielectronic recombination occurs. 
In this case, the Ar$^{+}$ region
penetrates inside the He$^+$ region, and the \exar/ is decreased (see
Sect.~\ref{sub:alphaAr} for the effects of dielectronic recombinations
of Ar$^{++}$.) Nevertheless,
using atmosphere models instead of BlackBody leads to a more important
increase of \exar/ while increasing \um/, as seen in
Fig.~\ref{fig:u_and_metal}. 

\begin{figure}
\epsfxsize=9.0cm  \epsfbox{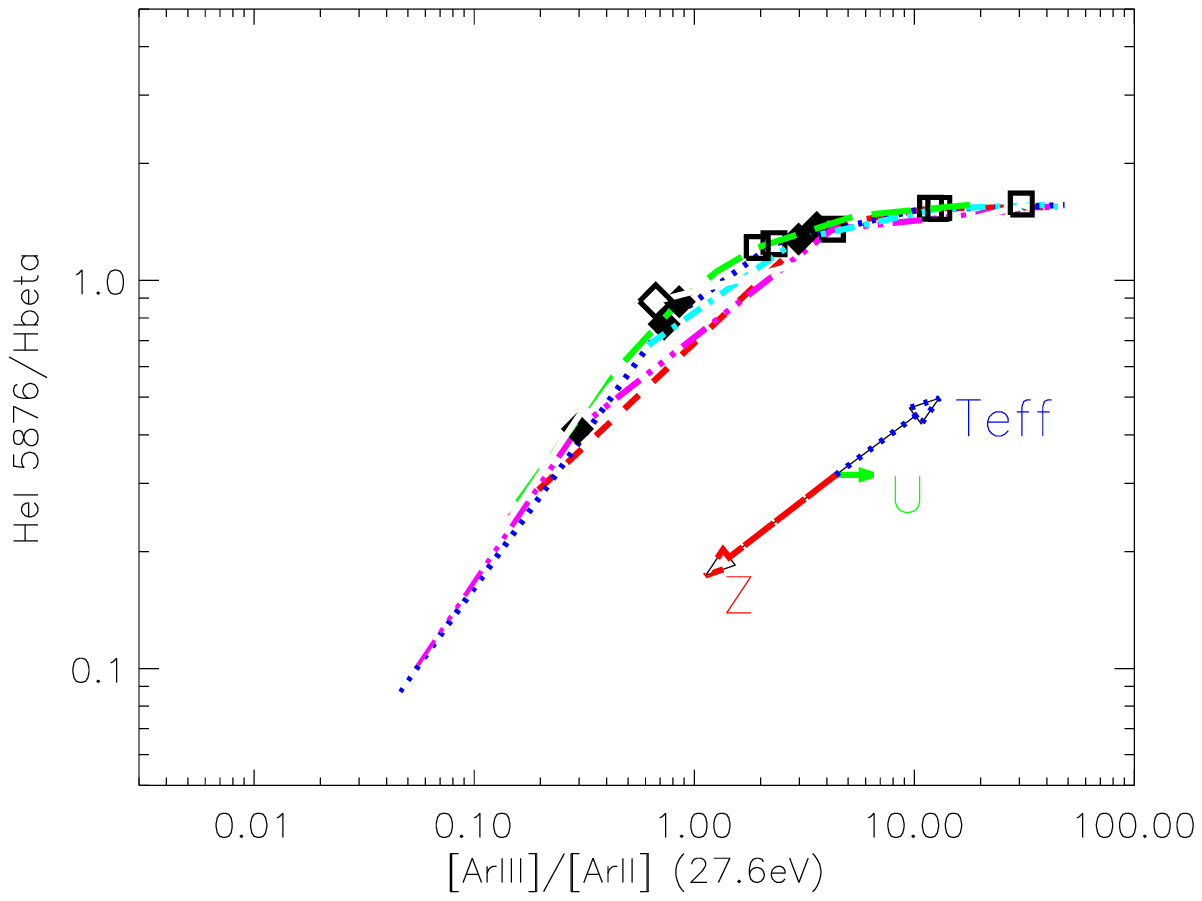}
\caption{Correlation between \heish/ and \exar/.
The line styles are the same as in Fig.~\ref{fig:comp_uv}. Arrows have the
same meaning as in Fig.~\ref{fig:maineffects}, for \wmbasic/ models only.
\label{fig:heishexar}}
\end{figure}

The extreme correlation between He$^+$/H and
Ar$^{++}$/Ar$^{+}$ can be verified in Fig.~\ref{fig:heishexar}, where
\heish/ versus \exar/
is plotted, for all the 
atmosphere models. While the \heish/ ratio saturate at a value
between 0.1 and 0.2, the \exar/ excitation diagnostic still evolve with
\teff/. This will be discussed further in Sect.~\ref{sub:teff}.

\subsection{Comparing results of Supergiant and Dwarf stars}
\label{sub:dwarfs}

\begin{figure}
\epsfxsize=9.0cm  \epsfbox{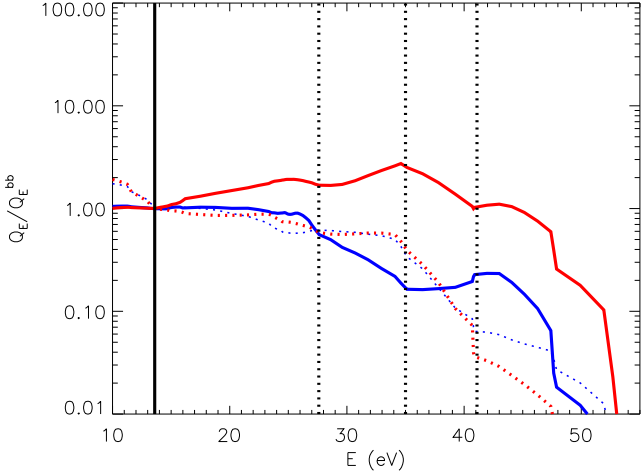}
\caption{Comparison between energy distribution (same as right panels of
Fig.~\ref{fig:comp_uv}) between Supergiants (light curves) and Dwarfs (bold
curves) of \wmbasic/ (dotted) and \cmfgen/ (dashed) models, all at 35~kK.
\label{fig:compSP_Naines}} 
\end{figure}

\begin{figure}
\epsfxsize=9.0cm  \epsfbox{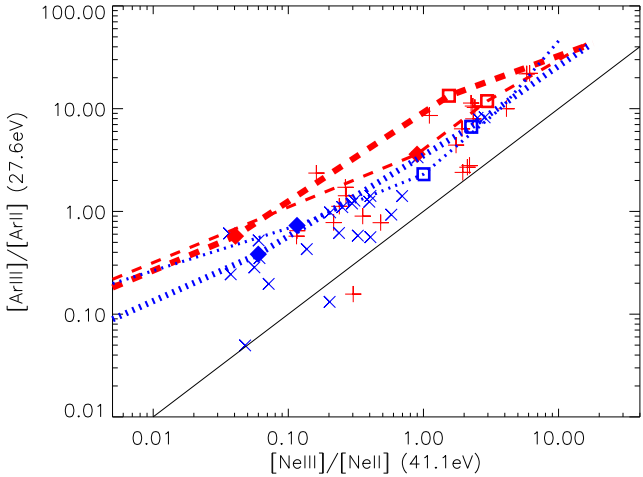}
\caption{Comparison between excitation diagnostic \exar/ obtained
with Supergiants (light 
curves) and Dwarfs stars (bold curves), for \wmbasic/ (dotted) and
\cmfgen/ (dashed) atmosphere models. Models at 35 and 40~kK are shown by
filled diamonds and empty squares respectively. \label{fig:compD}} 
\end{figure}

The results shown in previous Sections are obtained for Supergiant stars.
Models were also performed using Dwarfs stars (see description in
Sect.~\ref{sub:dwarfsdescr}). 
The decrease of \logg/ changes the shape of the ionizing radiation, as
shown in Fig.~\ref{fig:compSP_Naines}, where Q$_{\rm E}$ is shown for
Supergiants and Dwarfs atmosphere models obtained using \cmfgen/ and
\wmbasic/, all at 35~kK. The main effect of increasing \logg/, observed on the
two models, is to decrease strongly Q$_{\rm E}$ at 41~eV, and to increase the
slope of Q$_{\rm E}$ between 27 and 41~eV.
This overall hardening of the ionizing flux for stars with lower gravity
is due mostly to an increased ionization in the continuum forming layers, 
the latter effect resulting from the increased wind density (mass-loss rate).

Fig.~\ref{fig:compD} shows the excitation diagram \exne/ versus \exar/
performed using the models described above, comparing the Supergiant
(light curves) and Dwarf (bold curves) results for both \cmfgen/
(dashed) and \wmbasic/ (dotted).

As expected from the increased hardness of the ionizing fluxes for
supergiants,
the use of dwarf atmospheres leads in general to an excitation 
decrease which is more important at the highest energies 
(i.e.\ \exne/ decreases more rapidly 
than \exar/)\footnote{The \wmbasic/ model at 35 kK shown here is an
exception to this trend, however.}.
Overall the differences between supergiant and dwarf spectra do not
importantly affect our conclusions.

\subsection{Effect of changing the mean ionization parameter \um/}
\label{sub:u}

Given that the ionizing sources and the nebular geometry 
of the observed objects are essentially unknown and \um/ therefore
as well, it is important to examine how robust the above results
are with respect to changes of \um/.
E.g.\ is it possible to reconcile the discrepant predictions using 
the \kurucz/ and \tlusty/ atmosphere models (i.e.\ to increase the 
predicted \exne/ ratio of a given \exar/) by invoking
a larger ionization parameter toward the low excitation end
of the observed sequence?
As shown in Sect.~\ref{sub:maineffects}
and by detailed model calculations, variations of \um/
imply changes nearly parallel to the ``standard'' sequences
for the  \kurucz/ and \tlusty/ atmosphere models, similar to the case
of \wmbasic/ models shown in Fig.~\ref{fig:maineffects}.

We can therefore quite safely conclude that variations of \um/ 
cannot reconcile the \kurucz/ and \tlusty/ atmosphere models 
with the observation. 
However, for the other model atmospheres showing more ``curved'' 
predictions in the excitation diagrams, variations of \um/
might be invoked to improve/alter the predicted sequences
for constant \um/.

\subsection{Effect of adopting a metallicity gradient, for both the
stellar atmosphere and the ionized gas}
\label{sub:metal}

It is well
known that the metallicity decreases in the Galaxy when the distance to
the center increases \citep[e.g. ][and references
therein]{GMS02,PaperII}.
The metallicity varies approximately by a factor 4 from the center out
to 15~kpc, where the most external regions used in this work are
located (cf.\ Fig.~\ref{fig:exar_z}).

Fig.~\ref{fig:u_and_metal} shows the effect of changing the metallicity
$Z$ coherently in the ionizing star's SED computation and of the ionized
gas in the photoionization computation, from half 
solar to twice solar.
A metallicity increase in the atmospheres leads to a stronger
blanketing with the effect of softening the EUV spectra of early type
stars \citep[e.g.][]{PHL01}, leading thus to a lower nebular excitation.
As seen by comparing the Z-arrows in Fig.\ \ref{fig:u_and_metal} and by
additional test calculations, 
the increase of the nebular abundance plays a minor role in the
resulting excitation shift.

In fact the \wmbasic/ models used here show that the ionizing spectra
soften too strongly with increasing metallicity, leading to
a stronger reduction of \exne/ compared to \exar/. This results 
in a progressive shift away from the observed sequence toward
higher metallicity.
This discrepant trend, also found by \citet{GSL02}, was actually
used by these authors to argue that the observed excitation sequence
was mostly driven by \teff/ variations. However, as abundances
of these sources are known to vary by approximately the same factor
as the $Z$ variations considered here for the \wmbasic/ models,
metallicity cannot be neglected. Therefore the discrepancy between
the observations and the expected changes of \exne/ and \exar/ show
that the predicted softening of the \wmbasic/ ionizing spectra with metallicity
at high energies ($\ga$ 41 eV) is probably incorrect.
Alternate solutions to this puzzle include postulating a increase of
\um/ toward higher $Z$ (cf.\ above), or processes currently not
accounted for in the \wmbasic/ models altering the high energy part 
of the SED (cf.\ Sect.\ \ref{sub:not}), or changes in atomic physics
parameters (cf.~Sect~\ref{sub:alphaAr}).

\subsection{Effect of the dust}
\label{sub:dust}
The effect of the presence of dust inside an \hii/ region is firstly to
decrease the global amount of ionizing photons from the point of view
of the ionized gas, the effect being then to reduce the ionization
parameter. 
On the other hand, the efficiency of dust in absorbing of the ionizing
photons inside the 
\hii/ region decreases with the energy of the photons after about 18~eV
\citep[e.g.][]{M85,AA89},
the global effect being to increase the excitation of the gas when
increasing the amount of dust, for a given ionization parameter. 
As already pointed out by
\citet{PaperIII} for the case of ISO observations of G29.96-0.02, if
dust is present in \hii/ regions, quite the same  
excitation of the gas will be recovered using an higher ionization
parameter and a lower \teff/. As
illustration, inclusion of Graphite and 
Astronomical Silicate dusts, in proportion of 2.5\dix{-3}
relative to hydrogen, for each type of dust, leads to an increase of all the
excitation diagnostic ratios by a factor close to 2.
The excitation increase due to dust is found to be ``parallel'' to a \teff/ 
increase to reasonable accuracy, when keeping \um/ constant.

\subsection{Effect of the atomic data}
\label{sub:alphaAr}

Could uncertainties in the atomic data affect the results ? Indeed,
from Fig.~\ref{fig:exci1} to \ref{fig:exci3}, we could
suspect the Argon ionization equilibrium to be wrong, favoring the
emission of [\ion{Ar}{iii}]. This could e.g.\ be due to an overestimation
of the ionizing flux at 27.6~eV with respect to higher energies, 
or to a systematic error in the
observed intensities of one of the two lines involved in the \exar/
ratio (\forb{Ar}{iii}{8.98} is affected by silicate band).
However, we cannot exclude
also the effect 
of atomic data in the photoionization computation.
Collisional rates are generally believed to be
accurate within 20\%, while our knowledge of recombination
coefficients are less probant. Dielectronic recombination coefficients for the
elements of the third row of the periodic Table are poorly known, and
even the new computations done 
today are usually 
only for the first and second rows, corresponding to 
highly charged elements of the third rows, which is not the case for
Ar$^{++}$ \citep[see e.g.][]{SL02}. Dielectronic recombination
coefficients have been computed for less charged third row elements by
e.g. \citet{M98},
but only for high electron temperature (coronal gas), which is not the
case for \hii/ regions. Very recently, new dielectronic recombination rates
have been computed for Ar$^{++}$ (Zatsarinny et al., 2003, private
communication), but the results these authors obtain have still to be
checked (dielectronic recombination rates reach values as 10$^3$ times the
classical recombination rates for electron temperature closed to 10$^4$K!).

\begin{figure}
\epsfxsize=9.0cm  \epsfbox{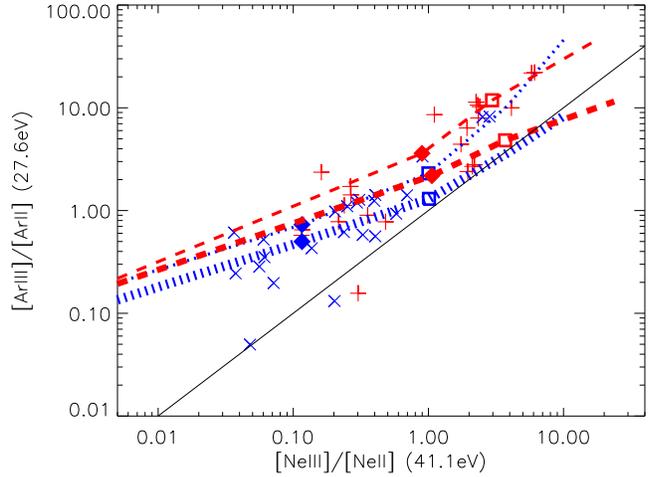}
\caption{Variation of the excitation diagnostics \exne/ versus \exar/
for the same 
atmosphere models (here \wmbasic/ and \cmfgen/, dotted and dashed
thin lines respectively), multiplying the effective
recombination coefficient for only Ar$^{++}$ by 10 (bold lines). 
\label{fig:alphaAr}}
\end{figure}

To simulate the effect of dielectronic recombination and charge
transfer reactions
we have multiplied the classical
recombination coefficient for Ar$^{++}$ by factors up to 20. 
Fig.~\ref{fig:alphaAr} shows
the effects of multiplying this coefficient arbitrarily by 10, on the
excitation diagnostics \exar/ versus \exne/. The figure shows results
using \wmbasic/ and \cmfgen/ models, but the same effect can be
observed with any of the atmosphere models used in this paper. 

The increase of the recombination coefficient improves overall the 
agreement with the observed sequence, although new discrepancies appear 
now at the high excitation end. 
However, no dependence of the dielectronic recombination coefficient 
on the electron temperature $T_e$ has been taken into account here.
As $T_e$ is known to vary along the excitation sequence this could
in fact ``twist'' the global shape of the predicted excitation sequence.
Currently both the exact ``direction'' and importance of this effect 
remain, however, unknown.
Note, that the uncertainties due to atomic data of Ar were already pointed
out by \citet{SSZ02}.
We join these authors in encouraging atomic physicists to improve
our knowledge of such data.  

\subsection{Effects not included in atmosphere models}
\label{sub:not}

What could be the limitations of the most sophisticated atmosphere
models currently available, capable of altering the excitation
diagnostics discussed here?
Although an exhaustive discussion is obviously not possible,
one can suspect one major process, namely the presence of X-rays,
to alter in a non-negligible way the ionizing spectra of O stars.
This has been shown clearly by \citet{Mac94}, and has been discussed
later e.g.\ by 
\citet{SC97}. The relative importance of X-rays compared to normal
photospheric emission is expected to increase for stars with
weaker winds and toward later spectral types. In late O types their
contribution can be non-negligible down to energies $\ga$ 30 eV, see
e.g.~\citet{Mac94} and a model at \teff/=30 kK by \citet{PHL01}, with
obvious consequences on nebular 
diagnostics.
Regrettably few models treating the X-ray emission in O stars exist, 
their impact on the overall emergent spectrum including the EUV
has hardly been studied with complete non-LTE codes including winds
and blanketing, and their dependence with stellar parameters
(wind density, stellar temperature, even metallicity?!) remains
basically unknown.

For now, we can only qualitatively expect the inclusion of X-rays to harden
the ionizing spectra, probably down to the energy range probed by
(some) mid-IR diagnostics. 
While this could in principle improve some difficulties observed 
by the \cmfgen/ and \wmbasic/ models (e.g.\ increasing \exne/)
their precise effect remains open.

\section{\teff/ diagnostics and ``second-order'' diagnostic diagrams}
\label{sec:diag}
\subsection{Using excitation diagnostic to determine \teff/}
\label{sub:teff}

\begin{figure}
\epsfxsize=9.0cm  \epsfbox{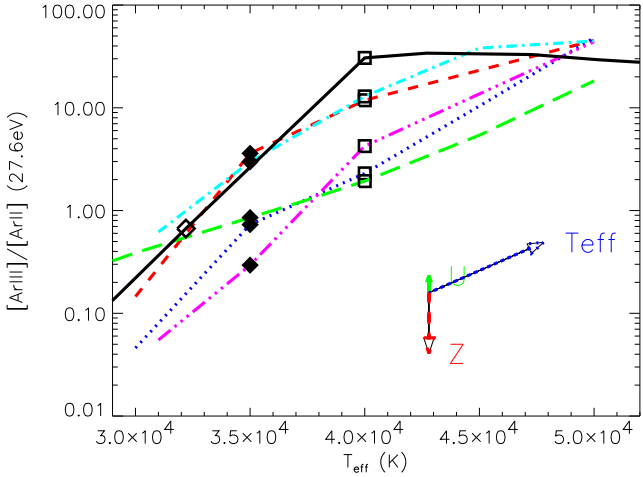}
\caption{Variation of the excitation diagnostic \exar/
according to the \teff/ of the ionizing star. Codes are
the same as in \figscode/. The set of arrows have the
same meaning as in Fig. \ref{fig:maineffects}, for \wmbasic/ models only.
\label{fig:t_diag}}
\end{figure}

\begin{figure}
\epsfxsize=9.0cm  \epsfbox{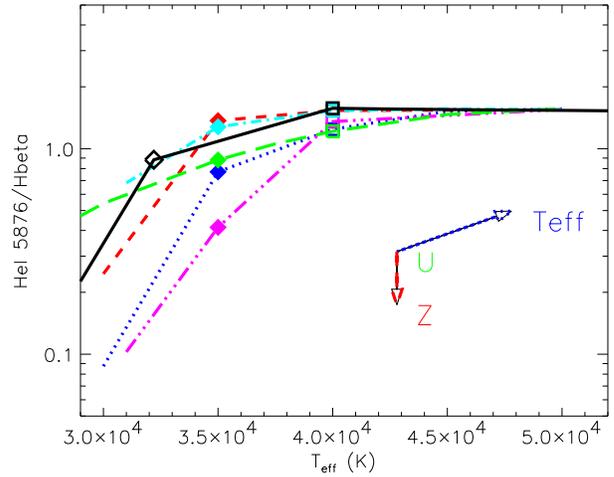}
\caption{Variation of the optical excitation diagnostic \heish/
with \teff/ for supergiant stars. Codes are
the same as in \figscode/. The set of arrows have the
same meaning as in Fig. \ref{fig:maineffects}, for \wmbasic/ models only.
\label{fig:t_diag_hei}}
\end{figure}

\begin{figure}
\epsfxsize=9.0cm  \epsfbox{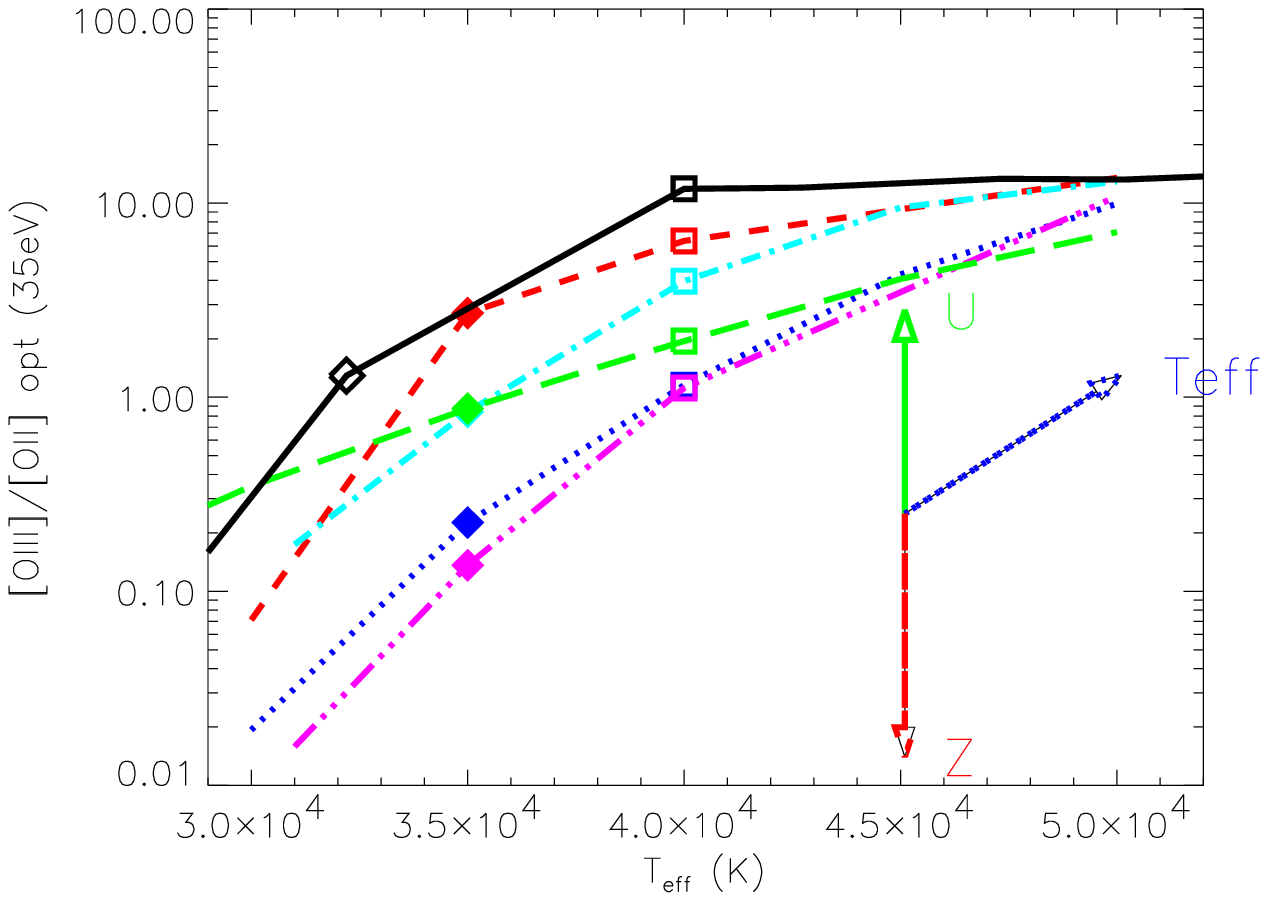}
\caption{Variation of the optical excitation diagnostic \exo/
with \teff/ for supergiant stars. Codes are
the same as in \figscode/. The set of arrows have the
same meaning as in Fig. \ref{fig:maineffects}, for \wmbasic/ models only.
\label{fig:exo_t}}
\end{figure}

Since the excitation of the gas increases with \teff/,
it is tempting to infer stellar temperatures from excitation 
diagnostic ratios. However, such an approach is intrinsically
highly uncertain, as the nebular excitation is also strongly
dependent on other parameters (see Sect.\ \ref{sec:grid}), 
such as the ionization parameter \um/, which remain in most cases
poorly known, cf. \citet{M82} for optical
lines and \citet{SS99} for mid-IR ratios. 
These cautionary remarks should be kept in mind when e.g.\
using single line ratios or even several line ratios 
\citep[e.g.][]{T00,Oka01,Oka03}, but see also \citet{M03},
to estimate stellar properties of individual objects from nebular observations.
Tailored photoionization models including numerous constraints
can lead to substantially different results and should clearly
be the preferred method \citep[see e.g.][]{PaperIII}.

For illustration we show in Fig.~\ref{fig:t_diag} the dependence of the 
\exar/ excitation ratio on \teff/ for a fixed \um/ and metallicity.
Other mid-IR excitation diagnostics show similar behaviors as can
already be seen from various figures above, except that their
dependence upon \um/ are higher than for \exar/, as already discussed
in Sect.~\ref{sub:heish}. The discussion of \teff/ and \um/
determinations using mid-IR excitation diagnostics and
the \hii/ regions metallicities is developed in \citet{M03}. 

The most important conclusion from Fig.~\ref{fig:t_diag} is the
important difference of the excitation of the gas ionized by \wmbasic/
(dotted line)
and \cmfgen/ (dashed line) stars, for the same \teff/ and \um/, even if
the two types of models are showing the same behavior in
Fig.~\ref{fig:exci1} to \ref{fig:exci3}. 
In other words, taking for example a value of 10. for \exar/, we can
determine a \teff/ of 39~kK using \cmfgen/ and a value
of 45~kK using \wmbasic/. This behavior can easily be understood when
comparing the Q$_{\rm E}$ distribution, as shown in
Fig.~\ref{fig:comp_uv} and discussed in Sect. \ref{sub:comp_uv}.

Two of the classical ways to constrain \teff/ from optical observations
are to use the \heish/ ratio \citep[e.g.][]{KBFM00} or the \exo/ ratio \citep{DP03}.
The predictions for \heish/ are shown in
  Fig.~\ref{fig:t_diag_hei}. 
As for \exar/, this line ratio is fairly independent of \um/.
  This diagnostic line ratio saturates above \teff/ $\ga$ 40~kK, when
  Helium is completely ionized to He$^+$.
  Note that even among \wmbasic/ and \cmfgen/, which treat very
  similar physics, some differences in this \teff/ indicator remain.
  Furthermore note that the predicted Q$_{24.6}$/Q$_{13.6}$ 
  and hence \heish/ vary
  non-negligibly between dwarfs and supergiants \citep[see
e.g.][]{PHL01,S02}. 
The predictions for \exo/ are presented in Fig.~\ref{fig:exo_t}. 
As for the \heish/ ratio, the results differ strongly from one
atmosphere model to another one. The Oxygen excitation diagnostic is also
strongly sensitive to \um/ and $Z$, the effect of
$Z$ being slightly more
important than found by \citet{DP03}, but these authors used \wmbasic/ dwarfs
atmosphere models.

We can conclude from the above examples that any determination of
\teff/ based on diagnostic ratios can be reliable only if the
metallicity of the star is coherently taken into account, and if \um/
is also determined at the same time, as shown in \citet{M03}.

\subsection{Radiation softness parameters $\eta$}
\label{sub:eta}

\begin{figure}
\epsfxsize=9.0cm  \epsfbox{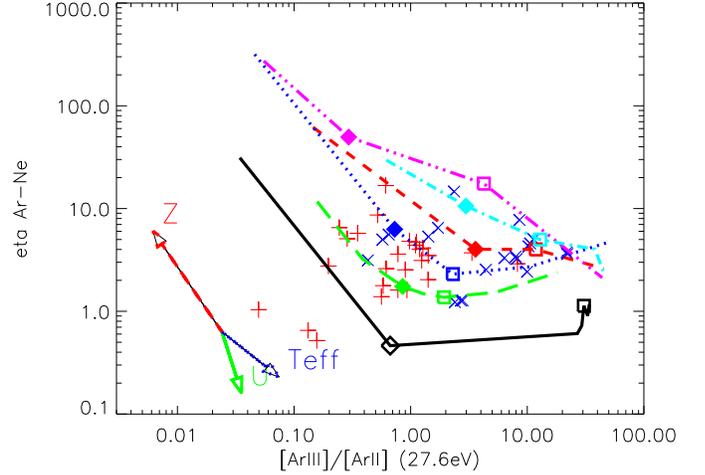}
\caption{Variation of the radiation softness parameter $\eta_{Ar-Ne}$
versus \exar/. The codes are the same than in \figscode/.
Note that the model predictions for both axis depend also on the
assumed ionization parameter \um/. Arrows as in
Fig. \ref{fig:maineffects} for \wmbasic/ models.
 \label{fig:eta_exci}}
\end{figure}

\begin{figure}
\epsfxsize=9.0cm  \epsfbox{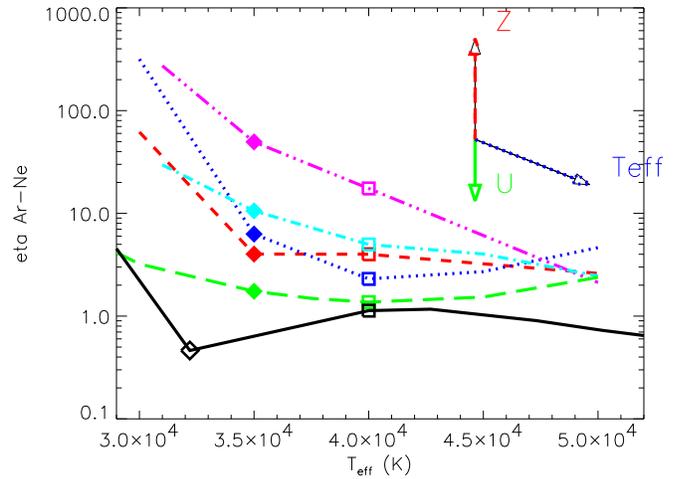}
\caption{Variation of the radiation softness parameter $\eta_{Ar-Ne}$
according to the \teff/ of the ionizing star. Codes are
the same as in \figscode/. Arrows have the same meaning as in
Fig.~\ref{fig:maineffects}, for \wmbasic/ models.
\label{fig:eta_diag}}
\end{figure}

\begin{figure}
\epsfxsize=9.0cm  \epsfbox{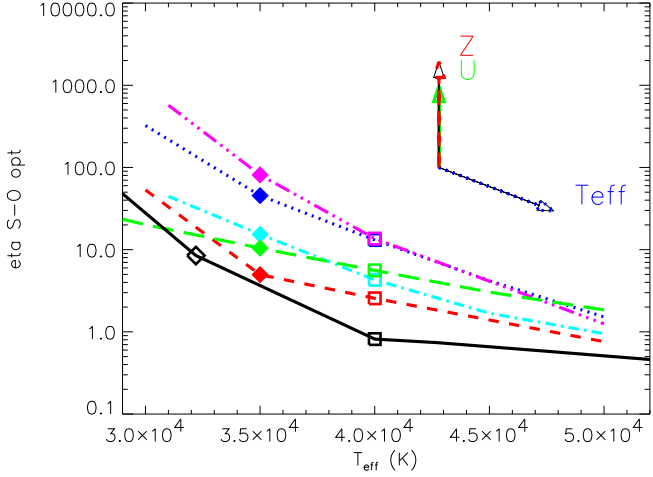}
\caption{Variation of the radiation softness parameter
$\eta_{S-O}$ (defined by \citet{VP88}, see note~\ref{not:etaopt}),
according to the \teff/ of the ionizing star. Codes are
the same as in \figscode/. 
Arrows have the same meaning as in Fig.~\ref{fig:maineffects}, for
\wmbasic/ models, for \wmbasic/ models only.
\label{fig:eta_diag_opt}}
\end{figure}

Following \citet{VP88}, radiation softness parameters can be defined
combining the excitation diagnostic ratios, namely: 
$\eta_{Ar-Ne} = \frac{\exar/}{\exne/}$
and so on for the other diagnostic ratios. 

To first order (but see the discussion on \exar/ in
Sect.~\ref{sub:maineffects}) an excitation ratio $X^{i+1}/X^i$ 
depends on the ionization parameter \um/ and the hardness of the
ionizing radiation and is given by

\begin{equation}
  \frac{X^{i+1}}{X^i} \propto \bar{U} \frac{\int_{\nu(X^i)}^\infty \frac{J_\nu}{h\nu} d\nu}
                                   {\int_{\nu(H^0)}^\infty \frac{J_\nu}{h\nu} d\nu}
                               = \bar{U} \frac{{\rm Q}_{E(X^{i})}}{{\rm Q}_{13.6}}.
\label{eq:xi}
\end{equation}

Therefore one has:
\begin{equation}
  \eta_{X-Y} \propto  \frac{{\rm Q}_{E(X^{i})}}{{\rm Q}_{E(Y^{i})}},
\label{eq:eta}
\end{equation}

where $E(X^{i})$ is the ionization potential of the ion $X^{i}$.
$\eta$ is thus in principle 
independent of the ionization parameter \um/, and a measure of the
``slope'' of the ionizing spectrum between the ionization energies
${E(X^{i})}$ and ${E(Y^{i})}$ respectively\footnote{However, note that
e.g.\ any $\eta$ involving \exar/ will likely depend on \um/, as \exar/
itself depends already little on \um/}.
Therefore such quantities -- used so far for optical lines only -- have 
often been thought to be good estimators
of stellar effective temperatures, see e.g. \citet{G89, KBFM00} 
but also \citet{S89,BKG99,OSDS02}.
For this reason we here explore whether the observed mid-IR fine structure
lines offer, both from the empirical and theoretical standpoint,
such a diagnostic power.

Fig. \ref{fig:eta_exci} shows the variation of $\eta_{Ne-Ar}$
versus \exar/. No correlation between these two observables is
found. While a galactic gradient is found for
\exar/ \citep{PaperII,GSL02}, $\eta_{Ne-Ar}$ 
and none of the other mid-IR $\eta$'s which can be constructed
show a galactic gradient. This can also be seen from the
correlations between the various excitation ratios plotted
by \citet{PaperII}, their Fig.~10.
Already this finding indicates empirically that mid-IR softness
parameters do not carry important information on the stellar 
ionizing sources.

How do the models compare with the observed softness parameters?
Given the considerable spread between photoionization models
using different stellar atmospheres and the various discrepancies
already found earlier, it is not 
surprising that overall a large spread is also found here 
(Fig.~\ref{fig:eta_exci}).
Compared to the observations the blackbody SED seems again to fit best. 
The \wmbasic/, \cmfgen/ and \costar/ models are marginally
compatible with the observations each one on one side, while the
\tlusty/ and \kurucz/ results are definitively far away from the observed
values.
Note, however, that both theoretical quantities plotted here (including $\eta$)
depend also on the ionization parameter.

In Fig.~\ref{fig:eta_diag} we illustrate theoretical predictions of mid-IR 
softness parameters as a function of \teff/ for the case of $\eta_{Ne-Ar}$.
Note that most of the model atmospheres predict that $\eta_{Ne-Ar}$ becomes
insensitive to \teff/ above a certain value, here of the order of \teff/ $\ga$ 
35000 K (the exact value also depending on \um/).
Similar dependencies on the adopted model atmosphere and ``saturation effects''
have also been found for the traditional optical softness parameter
\citep[cf.][]{OSDS02,KBFM00}.

\begin{figure}
\epsfxsize=9.0cm  \epsfbox{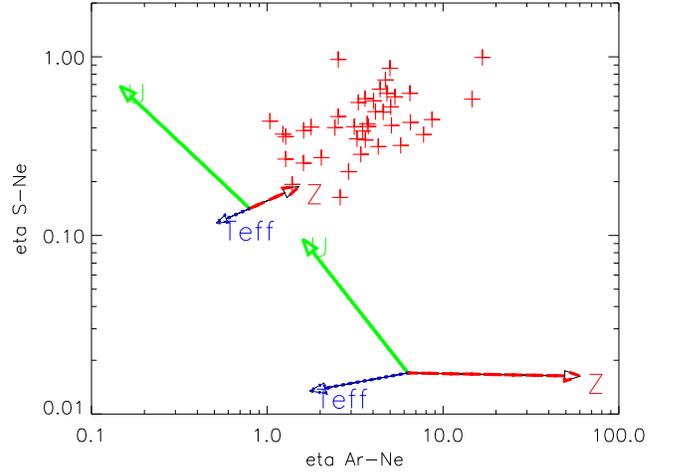}
\caption{Variation of the radiation softness parameter $\eta_{S-Ne}$
versus $\eta_{Ar-Ne}$, when changing \um/, \teff/ or Z. Arrows codes are
the same as in  Fig.~\ref{fig:u_and_metal}. \label{fig:eta_2_UZ}}
\end{figure}

Furthermore model calculations also show that all the mid-IR
$\eta$ depend quite strongly on the ionization parameter \um/
and on metallicity, as shown here for the case of the \wmbasic/
models\footnote{Note that the metallicity dependence predicted
by the \wmbasic/ models is probably overestimated as already 
discussed earlier (Sect.\ \ref{sub:metal}).} (see Fig.~\ref{fig:eta_2_UZ}).
From the theoretical point of view, and without tailored photoionization
modeling including constraints on \um/ and $Z$, the use of mid-IR
$\eta$'s appears therefore highly compromised. 
Again, similar difficulties have also been found for the optical
softness parameter, cf. \citet{S89,BKG99,OSDS02}, but \citet{KBFM00} 
and hereafter.

Last but not least, as each  $\eta$ implies 4 line intensities, 
the softness parameters are more sensitive to any observational uncertainty
(as the attenuation correction or detector calibration), 
possible collisional effects, uncertainties in the atomic data etc.
In view of all these considerations and the model results presented above,
applications of softness parameters in the mid-IR appear therefore 
to be of very limited use.

For completeness we show in Fig.~\ref{fig:eta_diag_opt} the behavior
of the traditional 
optical softness parameter $\eta_{O-S}$\footnote{Defined by \citet{VP88} as : 
(\forba{O}{ii}{3726,27}/\forba{O}{iii}{4959,5007}) / 
(\forba{S}{ii}{6717,31}/\forba{S}{iii}{9069,9532})\label{not:etaopt}}
for all model atmospheres. Again a considerable spread between different
atmosphere models and a dependence on \um/ and $Z$ is found.

            \section{Discussion}
\label{sec:disc}

From the comparisons of the extensive model calculations with the 
observed excitation diagnostics we can draw some rather general
conclusions on the shape of the ionizing spectra of early type stars
and the on the nature of the ionizing sources of the
Galactic \hii/ regions.

\subsection{Implications on stellar SEDs}
\label{sub:discsed}
Overall the ``best fit'' SED to the observed excitation diagrams
(Figs. \ref{fig:exci1} to \ref{fig:exci2}) was found with blackbody spectra. 
Does this imply that the ionizing fluxes of hot stars are best
described by the Planck function ?
The overall answer is no, but.
The observations probe (to 1st order) the relative number of
ionizing photons with energies above the relevant ionization
potentials, i.e.\ what was called the ``slope'' in Sect.~5.
Therefore if we considered that all observed 3 line ratios are 
correctly reproduced by a blackbody this would imply that
the 27.6--35.0 and 27.6--41.1~eV slopes (and hence also
35.0--41.1~eV) be equal to that of the Planck function (bb) of the
same \teff/, i.e.\
${\rm Q}_{27.6}/{\rm Q}_{35.0}=
{\rm Q}_{27.6}^{\rm bb}/{\rm Q}_{35.0}^{\rm bb}$
and 
${\rm Q}_{27.6}/{\rm Q}_{41.1}=
{\rm Q}_{27.6}^{\rm bb}/{\rm Q}_{41.1}^{\rm bb}$.
This would therefore represent three ``integral'' constraints on 
the stellar SED. This is rather strong, but still leaves room for the 
detailed shape of the SED between these energies.
Actually the agreement between blackbody spectra and \exar/ vs.\
\exne/ is better than diagrams involving \exs/.
This indicates that the constraint on the ${\rm Q}_{27.6}/{\rm Q}_{41.1}$
is better than at intermediate energies.

However, it is important to remember that these
constraints on the underlying SED can be deduced only if 
the observed excitation diagram (Figs.\ref{fig:exci1} to \ref{fig:exci3}) 
is essentially driven by a temperature sequence (i.e.\
$\bar{U}=const$). Note also the effects of uncertainties on the atomic
data, as discussed in Sect.~\ref{sub:alphaAr}.

\subsection{On the origin of the observed excitation sequences}
\label{sub:origin}

\begin{figure}
\epsfxsize=9.0cm  \epsfbox{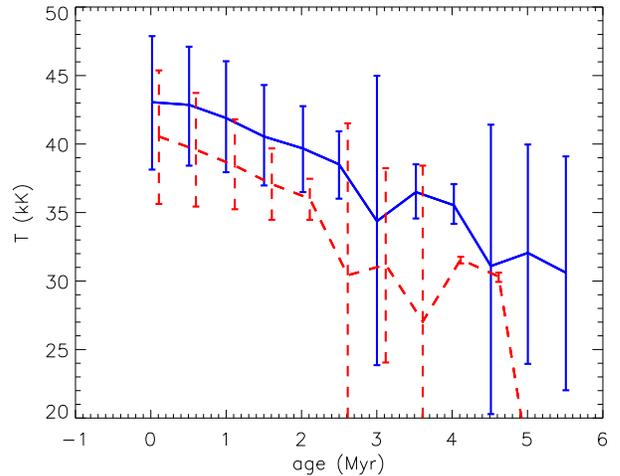}
\caption{Average stellar effective temperature and dispersion 
as a function of time for metallicities $Z=0.008$ ($1/2.5 Z_\odot$,
solid lines) and 0.04 ($2  Z_\odot$, dashed) predicted for an
ensemble of single star \hii/ regions for a Salpeter IMF with 
$M_{\rm up}=120 M_\odot$ and a minimum Lyman continuum flux
$\log({\rm Q}_{\rm lim}) \ge 48.$.
Note the rather small differences in average \teff/ and the large
intrinsic dispersion for each given metallicity.
\label{fig:mc_teff}}
\end{figure}

The origin of the observed mid-IR excitation sequences 
and their correlation with galactocentric distance has been 
discussed recently by \citet{GSL02,MVT02}. 
As both studies present somewhat limited arguments a
more general discussion is appropriate here.
The basic question is whether variations of the stellar effective
temperature or metallicity variations are responsible for the 
observed decrease of excitation toward the Galactic Center?

\citet{GSL02} have noted from photoionization models using
\wmbasic/ spectra that \exne/ is predicted to decrease more 
rapidly than \exar/ with increasing metallicity -- 
a finding also confirmed here (cf.\ Fig.~\ref{fig:u_and_metal}). 
However, as the observed sequence does NOT
follow this trend, they conclude that the decrease of excitation
must be due to a reduction of \teff/ as opposed to a softening
of the stellar SED with increasing metallicity.
Obviously as such this conclusion cannot be upheld as the same mid-IR observations,
allowing fairly accurate abundance determinations, clearly establish
the existence of a metallicity gradient \citep{GMS02}.
In fact, the apparent contradiction between the observed trend with metallicity
and the one predicted with \wmbasic/ model atmospheres indicates quite
likely that the stellar SEDs soften too quickly with increasing metallicity
and/or that the ionization parameter in regions at small galactocentric
distance must be larger than assumed (cf.\ Sect.~\ref{sub:metal}).

In contrast to the above study, \citet{MVT02} show a loose correlation
between excitation and metal abundances (e.g.\ between \exne/ and Ne/H),
stress the importance of metallicity effects on the SED \citep[see also][]{MH03}, 
and conclude that at least partly the observed decrease
of \exne/ must be due to a softening of the stellar SEDs with increasing
metallicity.

\subsubsection{Stellar evolution effects}

From what we know, three effects are related to metallicity
and must all be taken into account. First, higher metallicity is known
in stellar evolution to lead to a cooler zero age main sequence and 
to an overall shift to cooler temperatures \citep[e.g.][]{SSMM92}.
Second, blanketing effects in the atmospheres become stronger with 
increasing metallicity and lead to softer ionizing spectra
\citep[e.g. Sect.~\ref{sub:metal};][]{PHL01,MH03}.
Finally, an increased nebular abundance leads also to a somewhat
lower excitation of the gas in the \hii/ regions (cf.~Sect.~\ref{sub:metal}).
The remaining questions are then {\em a)} which of these effects dominate
and {\em b)} whether taken together they can indeed quantitatively reproduce 
the entire range of observed excitation variations in the Galactic \hii/ regions.
\begin{figure}
\epsfxsize=9.0cm  \epsfbox{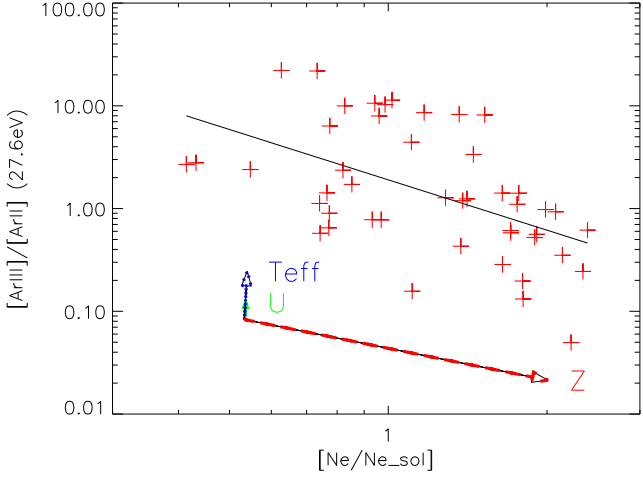}
\caption{\exar/ versus 
metallicity, measured here by
the Ne abundance (see text).
The line is a linear fit to the
observations in log-log 
space. Arrows as in Fig.~\ref{fig:maineffects}, for \wmbasic/ models.
 \label{fig:exar_z}}
\end{figure}
\begin{figure}
\epsfxsize=9.0cm  \epsfbox{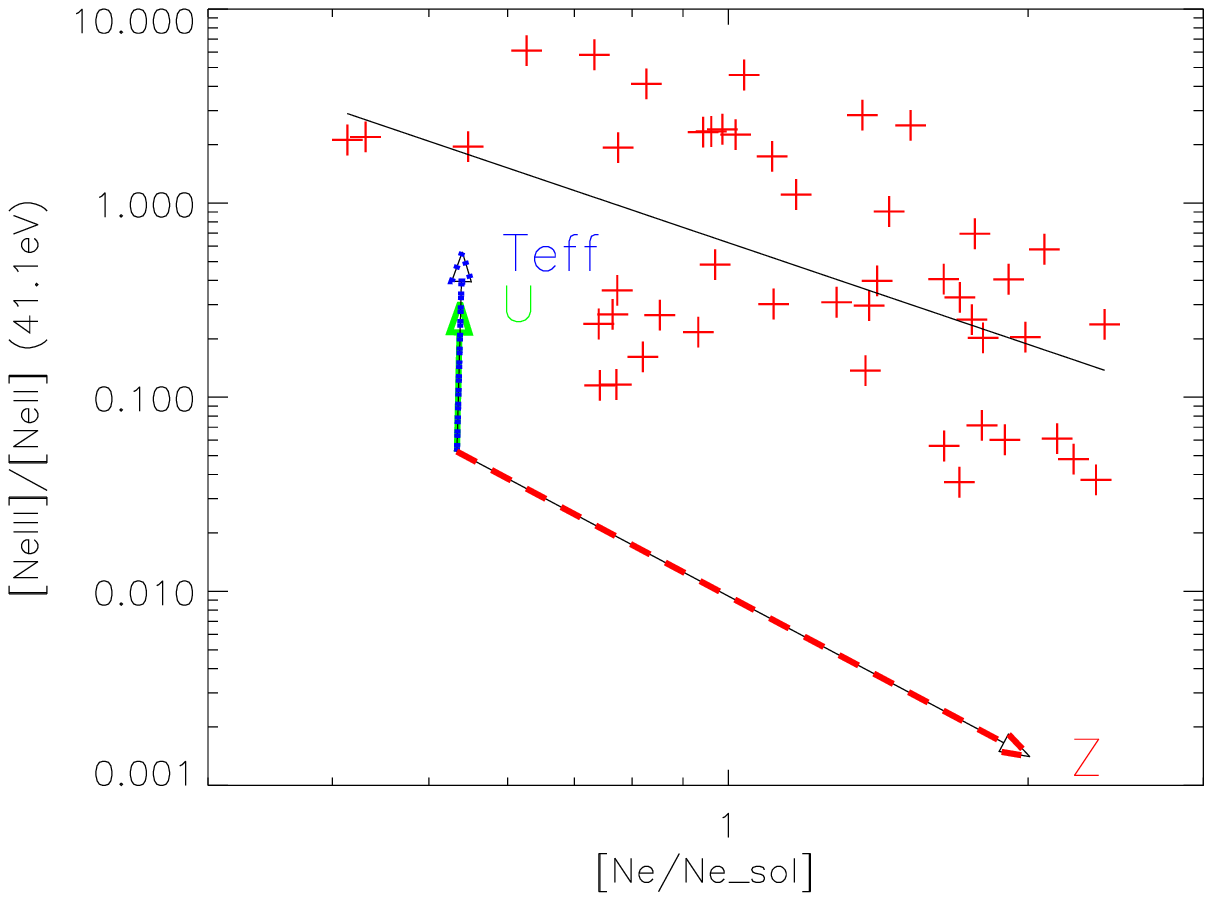}
\caption{\exne/ versus metallicity, as in
Fig.~\ref{fig:exar_z}. A very similar plot is obtained for \exs/.
 \label{fig:exne_z}}
\end{figure}

Although the content of stellar ionizing sources of the objects considered
is not known \citep[but see][]{M03} we can estimate the \teff/ variations
expected from 
stellar evolution, e.g.\ by assuming a single ionizing source. 
We then perform Monte Carlo simulations of single star \hii/ regions
of different metallicities assuming that the ionizing stars
are drawn from a Salpeter IMF with a given upper mass cut-off
$M_{\rm up}$.
In order to compute the mean properties of these stars, such as their
average \teff/, the predicted variation with metallicity and their
dispersion, the equivalent of a lower mass limit must also be specified. 
This is done by imposing a lower limit on the total Lyman continuum photon
flux ${\rm Q}_{\rm lim}$. Only stars with $Q >{\rm Q}_{\rm lim}$ at a given
age are retained for this computation.  
In practice we use the \citet{MMS94} stellar tracks, we consider 
metallicities between $\sim 1/2$ and 2 times solar, as indicated
by the observed range of Ne/H or Ar/H abundances and we adopt
$\log({\rm Q}_{\rm lim})=48.$, corresponding a typical lower limit
for the \hii/ regions of \citet{PaperII}.
Very massive stars entering the Wolf-Rayet phase already on the main sequence
are also excluded.

The resulting average and spread of \teff/ as a function of age is shown
in Fig.~\ref{fig:mc_teff} for $Z=0.04$ and 0.008 respectively.
This figure shows the following:
first, the reduction of the average stellar temperature due to a metallicity
increase by a factor 4 is rather modest, of the order of $\sim$ 3-4 kK.
Second, for a given metallicity the predicted dispersion in \teff/ is
larger; typically of the order of $\sim$ 3-9 kK for reasonable ages.
Although the absolute values of these \teff/ depend for obvious
reasons on the exact choice of ${\rm Q}_{\rm lim}$, the \teff/ differences
and dispersion depend little on this value.
From all the models considered above, the decrease of the mean \teff/
due to stellar evolution effects appears to be too small to explain 
the full range of excitations.

On the other hand the fairly large \teff/ dispersion at a given $Z$
will induce  
quite important variations in the excitation. This effect probably
dominates the observed excitation scatter (defined at the end of
Sect.~\ref{sec:exci}), as also suggested by the 
observations of a large spread of the excitation diagnostics 
Figs.~\ref{fig:exar_z} and \ref{fig:exne_z} for a
given metallicity measured here by Ne/Ne$_\odot$
\footnote{ Ne abundances were obtained using reddening and 
T$_{\rm elec}$ corrected abundances from \citet{GMS02}. 
Ne is preferred to Ar or S because Ar 
abundance is not reliable for high 
\teff/, Ar$^{3+}$ being present but unobserved, the same applying
at low \teff/ for S$^+$.
The solar abundance for Ne is determined
using the abundance gradient obtained by \citet{GMS02} at
8.5~kpc.}.

Indeed 
the observed spread of \exar/ 
and \exne/ at a given $Z$ is somewhat larger than or
similar to the decrease of the mean excitation with increasing $Z$.
Taken together these findings imply that while undeniably metallicity
effects on stellar evolution and nebular
abundances must be present, statistical fluctuations of the effective
stellar temperature due to the IMF are likely the dominant source of
scatter for the observed mid-IR excitation sequence of Galactic \hii/
regions, while the excitation sequences must be predominantly driven
by other effects which we will discuss now.

\subsubsection{Stellar atmospheres and nebular effects}
As apparent from our modeling (see \ Figs.\ \ref{fig:exar_z} and 
\ref{fig:exne_z}) the effects of metallicity on the shape of 
the stellar ionizing spectra strongly alter the predicted excitation.
The magnitude of the predicted effect is found to be comparable
to the observed variation.
Both these findings and the above results concerning stellar
evolution effects indicate the $Z$ dependence of the ionizing spectra
is the main driver for the correlation of the excitation with
galactocentric distance.

As this result depends on a specific set of model atmospheres
(\wmbasic/) a few words of caution are, however, necessary here.
First, we note that the effects of \teff/ and $Z$ are not 
exactly the same for \exar/,\exne/, and \exs/.
The predicted excitation variation (shown here for a change
of \teff/ from 35 to 40~kK)
tends to be somewhat larger (smaller) for \exne/ and \exs/ (\exar/) 
than the observed variations.
Second, it must be remembered that the \wmbasic/ atmosphere models
employed here could predict too strong a softening with increasing $Z$
as suspected from Fig.~\ref{fig:exne_z} and already discussed in
Sect.~\ref{sub:metal}.
Despite these imperfections there is little doubt that the
above result remains valid.

Finally we may also comment on excitation changes related to 
the ionization parameter.
Again, as for \teff/,  the above Monte Carlo simulations show small
differences between the average ionizing photon flux $Q_{13.6}$ 
with metallicity, but a considerable dispersion for each $Z$.
Combined with the observational fact of fairly similar nebular
densities in our objects this could be a justification for a constant
ionization parameter, at least on average.
Random variations of \um/ are, however, expected to contribute to
the excitation scatter at a given metallicity.

In conclusion we see little doubt that the observed excitation sequence of
Galactic \hii/ regions 
is shaped by the joint effects of metallicity on stellar evolution, 
atmospheric line blanketing, and cooling of the ISM.
From our investigations it seems, however, that metallicity effects on
ionizing stellar flux is the dominant effect causing the excitation
gradients while statistical
fluctuations of \teff/ and \um/ are likely the dominant source of
scatter in the observed excitations.
A more detailed study of possible systematic \teff/ and ionization
parameter gradients is presented in \citet{M03}: no clear gradient of
\teff/ nor \um/ versus the galactocentric distance are found by this
author, but some trends of increase (decrease) of \teff/ (\um/) with
the metallicity are observed. Importance of taking into account the
effect of the metallicity on the stellar spectral shape is addressed.

            \section{Summary and Conclusion}
\label{sec:concl}

We have presented results from extensive photoionization model grids
for single star \hii/ regions using a variety of recent
state-of-the-art stellar atmosphere models such as 
\cmfgen/, \wmbasic/, \tlusty/, \costar/, and \kurucz/ models.
In fact, even among the two recent non-LTE line blanketed codes
including stellar winds (\wmbasic/ and \cmfgen/) the predicted
ionizing spectra differ by amounts leading to observable 
differences in nebular spectra\footnote{
Presumably these differences are due to the use of different
methods to treat line blanketing (opacity sampling method versus
super-level approach in the comoving frame) and different atomic data.}

The main aim of this investigation was to confront these model
predictions to recent catalogs of ISO mid-IR observations of Galactic \hii/
regions, which present rich spectra probing the ionizing spectrum
between $\sim$ 24 to 41 eV thanks to the measurements of
\exar/, \exne/, and \exs/ line ratios.
Particular care has been paid to examining in detail the 
dependences of the nebular properties on the numerous nebular
parameters (ionization parameter \um/, abundances, dust etc.) which
are generally unconstrained for the objects considered here.
 
Most excitation diagnostics are found to be fairly degenerate,
but not completely so, with respect to increases of \teff/, \um/, a change
from dwarf to supergiant spectra, a decrease of the nebular
metallicity (Sects.~\ref{sub:maineffects} and \ref{sub:dwarfs}), and
the presence of dust in the \hii/ region (Sect. \ref{sub:dust}).
Each of these parameters increases the overall excitation of the gas, 
and in absence of constraints on them, a derivation of such a
parameter, e.g.\ an estimate of the stellar \teff/ of the ionizing 
source, is intrinsically uncertain.
In consequence,
while for sets of objects with similar gas properties statistical
inferences are probably meaningful, such estimates for individual
objects must be taken with care.

Provided the ionization parameter is fairly constant on average
and the atomic data is correct (but cf.\ below) the comparisons 
between the photoionization model predictions and the observations
allow us to conclude the following concerning the different
stellar atmosphere models (Sect.\ \ref{sub:scomp_obs}):

\begin{itemize}
\item
Both recent non-LTE codes including line blanketing and stellar winds
(\wmbasic/ and \cmfgen/) show a reasonable agreement with the observations.
Given their different behavior in the three excitation diagnostics,
depending on luminosity class, 
and other remaining uncertainties, it appears that none of the models
can be preferred on this basis.

\item 
The plane parallel hydrostatic codes (\kurucz/, \tlusty/)
predict spectra which are too soft, especially over the energy
range between 27.6, 35.0, and 41.1~eV and above.
Although a good agreement is found for UV to optical spectra
predicted by the hydrostatic \tlusty/ code and the photosphere-wind 
code \cmfgen/ \citep{H03,JCB03}
important differences are found in the EUV range probed by the
present observations and photoionization models.
Apparently the full non-LTE treatment of numerous elements
accounted for by \tlusty/ is insufficient to accurately predict the 
ionizing spectra at these energies, and the inclusion of
stellar winds is imperative.

\item
We confirm the finding of earlier investigations \citep[e.g.][]{ODSS00}
showing that the \costar/ models overpredict somewhat the ionizing 
flux at high energies.

\item Interestingly blackbodies reproduce best the observed
excitation diagrams, which indicates that the ionizing spectra of our
observed object should have relative ionizing photon flux productions 
${\rm Q}_{\rm E}$ at energies 27.6, 35.0 and 41.1~eV close to 
that of blackbody spectra.
Although this integral constraint on the SED remains approximate,
it should still be useful to guide future improvements in atmosphere 
modeling.

\item Finally, the softening of the ionizing spectra with increasing
metallicity predicted by the \wmbasic/ models is found to be too
strong. As already apparent from observed correlations between
excitation diagnostics probing various energies, the observed softening
of the radiation field (in part due to metallicity) affects fairly
equally the range between $\sim$ 27 and 41 eV \citep{PaperII} in
contrast to the atmosphere model predictions, which soften
most at the highest energies.
\end{itemize}

These conclusions are found to be fairly robust to effects such as
changes of \um/, nebular and stellar metallicity changes, and the inclusion
of dust.
We suggest that the main uncertainty which could alter the above
conclusions is the poorly known atomic data for Ar$^{++}$
(especially dielectronic recombination coefficients)
as also pointed out by \citet{SSZ02}. Reliable computations for such data
are strongly needed.
From the perspective of atmosphere codes probably the most important
step toward improving the reliability of ionizing fluxes
resides in a quantitative exploration of the influence of X-rays
on the emergent spectra at lower energy.

The potential of mid-IR line ratios or ``softness parameters'',
defined in analogy to the well known $\eta$ parameter for
optical emission lines, has been explored (Sect.\ \ref{sub:teff}).
The following main results have been obtained:

\begin{itemize}
\item
Given the non-negligible differences between the various atmosphere models
it is not surprising that individual line ratios (e.g.\ \exar/, \exne/)
show quite different dependences on \teff/.
We find that \exar/ depends little on the ionization parameter
as the ionization of Ar$^{+}$ is closely coupled to that of He.
This suggests that \forb{Ar}{iii}{8.98}/\forb{Ar}{ii}{6.98} 
should in principle be a fairly robust temperature indicator, 
provided the atmosphere models are sufficiently accurate
up to $\sim$ 24-27 eV and the atomic data is reliable
(cf.\ above). 

In comparison to \ion{He}{i}/H indicators 
\citep[e.g. \alloa{He}{i}{6678}/H$\beta$ or 
\allo{He}{i}{2.06}/Br$\gamma$ in][and references therein]{KBFM00,
Lum03}, \exar/ does not show a saturation effect but remain sensitive to
\teff/ up to the highest temperature examined here (\teff/ $\sim$
50~kK). Due to the large uncertainties of dielectronic recombination
coefficient of Ar$^{++}$, \citet{M03} prefer to use \exs/ and \exne/
to determine \teff/ and \um/ simultaneously, for  the \hii/ regions
used in this work. 

\item
Both empirically and theoretically the mid-IR softness parameters
which can be constructed from \exar/, \exs/, and \exne/ are found
to provide little if any information on stellar temperatures if not
used to determine \um/ at the same time \citep{M03}.
Observationally little / no correlation is found between the
$\eta$'s and excitation ratios.
Furthermore, our photoionization models show a considerable dependence
of $\eta$ on the ionization parameter.
We therefore conclude that mid-IR $\eta$'s 
appear to be of limited diagnostic power.
\end{itemize}

Finally we have examined which parameter(s) is (are) chiefly responsible
for the observed mid-IR excitation sequences.
Combining the results from our extensive photoionization model grids
with Monte Carlo simulations of ensembles of single star \hii/ regions
of different metallicity and age we conclude the following
(Sect.\ \ref{sec:disc}).
While metallicity effects on stellar evolution, atmospheres and the nebulae
all have an undeniable influence, they are probably of minor importance 
compared to the fairly large dispersion of \teff/ expected at each
metallicity from a simple statistical sampling of the IMF.
The \teff/ scatter plus additional scatter in the ionization parameter
are probably the dominant driver for the 
observed mid-IR excitation scatter of Galactic \hii/ regions, while
the effect of metallicity on the shape of the ionizing spectra is
partially responsible of the global excitation sequence, the proportion
of this effect being strongly dependent of the reliability of the
atmosphere models \citep{M03}.


\acknowledgements
We wish to thank various persons who have contributed directly
and indirectly to the shape of this paper, over its
rather long gestation time. Among those are in particular
Grazyna Stasi\'nska and 
Daniel P\'equignot who have been consulted on photoionization codes and
atomic physics, and Leticia Mart{\'\i}n-Hernandez.
Thierry Lanz kindly provided model atmosphere results prior
to publication. We are also grateful to John Hillier for assistance in
adapting and modifying his atmosphere code to our purpose.
We thank Ryszard Szczerba for helping introducing dust in NEBU code.
DS thanks the CNRS, the French Programme National de Galaxies, 
and the Swiss National Fund for Research for support.
DS and FM thank the CALMIP and IDRIS centers for generous
allocation of computing time.
CM thanks the IA/UANM in Mexico for offering him the opportunity to
finish this work and investigate many others in the future.
This work was partly supported by the CONACyT (M\'exico) grant 40096-F. 


\end{document}